\documentclass[aps,twocolumn,amsmath,floatfix]{revtex4-2}
\usepackage{lineno,isotope,mathrsfs}
\modulolinenumbers[5]
\usepackage{amsmath,bbold,amssymb,epsfig,bm,mathrsfs,feynmp}
\usepackage{color,slashed,exscale,multirow,soul}
\usepackage{times,txfonts,xcolor,graphicx,tikz}
\usepackage[normalem]{ulem}
\usepackage[breaklinks,colorlinks,citecolor=blue]{hyperref}
\definecolor{red}{rgb}{0.8,0,0}
\definecolor{violet}{rgb}{0.4,0,0.4}
\definecolor{green}{rgb}{0,0.5,0.0}
\definecolor{navy}{rgb}{0.0,0.0,0.6}
\definecolor{orange}{rgb}{0.8,0.2,0.0}
\usepackage[normalem]{ulem} %\sout{old text} for strikeout  

\begin{document}
\title{Universal relations for compact stars with heavy baryons}
\author{Jia Jie Li}
\email{jiajieli@swu.edu.cn}
\affiliation{School of Physical Science and Technology, 
Southwest University, Chongqing 400715, China}
\author{Armen Sedrakian}
\email{sedrakian@fias.uni-frankfurt.de}
\affiliation{Frankfurt Institute for Advanced Studies,
D-60438 Frankfurt am Main, Germany}
\affiliation{Institute of Theoretical Physics,
University of Wroclaw, 50-204 Wroclaw, Poland}
\author{Fridolin Weber}
\email{fweber@sdsu.edu}
\affiliation{Department of Physics, San Diego State University,
5500 Campanile Drive, San Diego, California 92182, USA}
\affiliation{Center for Astrophysics and Space Sciences,
University of California at San Diego, La Jolla, California 92093, USA}
\begin{abstract}
  A set of hadronic equations of state derived from covariant density
  functional theory and constrained by terrestrial experiments, and
  astrophysical observations, in particular by the NICER experiment
  inferences is used to explore the universal relations among the
  global properties of compact stars containing heavy baryons at high
  densities. We confirm the validity of universal $I$-Love-$Q$
  relations connecting the moment of inertia $(I)$, the tidal deformability 
  ($\Lambda$), and the spin-induced quadrupole moment ($Q$) for isolated
  non-rotating stars. We further confirm the validity of the
  $I$-$C$-$Q$ relations connecting the moment of inertia, compactness
  $(C)$, and quadrupole moment for uniformly and slowly rotating
  stars, and extend the validity of these relations to maximally
  rotating sequences. We then investigate the relations between
  integral parameters of maximally rotating and static compact
  stars. The universalities are shown to persist for equations of
  state and compositions containing hyperons and $\Delta$ degrees of
  freedom. When heavy baryons are included, however, the radial profiles 
  of integrands in expressions of global properties exhibit ``bumps", which 
  are not present in the case of nucleonic stars in which case the profiles 
  are smooth. We determine the coefficients entering the universal relations 
  in the case of hyperonic and $\Delta$-resonance containing stars.
\end{abstract}
\maketitle
%
%-------------------------------------------------------------
\section{Introduction}
\label{sec:Intro}
%-------------------------------------------------------------
Compact stars (CSs) as the densest objects in the observable universe
form natural astrophysical laboratories for understanding the physics of 
matter at supranuclear
densities~\citep{Lattimer:2012,Chatterjee:2015,Lattimer:2015,Oertel:2016,Ozel:2016,Baym:2018,Sedrakian:2018,Baiotti:2019,Tolos:2020,Logoteta:2021,Burgio:2021,Lattimer:2021,Sedrakian:2022ppnp}.
Astronomical observations of CSs impose constraints on the behavior of 
the equation of state (EoS) of dense matter, which is an important input 
for determining the macroscopic properties of CSs, such as the mass, 
radius, moment of inertia, etc. The groundbreaking detection of the first 
gravitational wave (GW) signals from a binary neutron star merger event 
GW170817~\citep{GW170817,LVC:2017} has opened a new avenue for studying 
the internal structure of CSs and the properties of dense stellar
matter~\citep{Annala:2017,Fattoyev:2017,Bauswein:2017,Most:2018,DeSoumi:2018,Coughlin:2018,Lijj:2019a}.
The X-ray pulse profile modeling of pulsars~\cite{Psaltis:2013} combined 
with inferences from the NICER 
experiment~\citep{NICER:2019a,NICER:2019b,NICER:2021a,NICER:2021b} 
led to measurements of CSs' masses and radii. The mass-radius ranges
derived for the two-solar mass PSR J0740+6620 are of great value for
inferring the properties of dense matter at sufficiently high
densities~\citep{Fasano:2019,Legred:2021,Raaijmakers:2021,Tangsp:2021,Zhangnb:2021,Lijj:2021,Christian:2021}.

The integral parameters of CSs such as the mass, radius, moment of
inertia, quadrupole moment, etc., are controlled by the microscopic
EoS of dense matter. Nevertheless, various approximately universal
relations connecting different CS integral parameters have been
established and intensively studied in recent
years~\citep{Yagi:2013a,Yagi:2013b,Doneva:2013,Maselli:2013,Baubock:2013,Haskell:2013,Haskell:2013,Pappas:2014,Yagi:2014,Chakrabarti:2014,Martinon:2014,Steiner:2015,Cipolletta:2015,Breu:2016,Staykov:2016,Yagi:2016,Doneva:2016,Paschalidis:2017,Silva:2017,Gagnon-Bischoff:2017,Marques:2017,Weijb:2018,Lenka:2018,Lim:2018,Stone:2019,Riahi:2019,Kumar:2019,Wendh:2019,Jiangnan:2020,Sunwj:2020,Raduta:2020,Koliogiannis:2020,Godzieba:2020,Nedora:2021,Khadkikar:2021,Saes:2021,Annala:2021,Danchev:2021,Kuan:2022,Konstantinou:2022,Wangsb:2022,Largani:2022,Zhaotq:2022}.
Because these relations are insensitive to the input EoS and are held to a
high accuracy (the typical deviations are at the level of several per
cent), they are called {\it universal}. The $I$-Love-$Q$ relations,
which connect the moment of inertia $I$, tidal deformability
$\Lambda$, and the spin-induced quadrupole moment $Q$ of CSs in slow
rotation approximation were first discovered in
Refs.~\citep{Yagi:2013a,Yagi:2013b}. These sorts of relations 
among various integral parameters of CSs have been studied
under various conditions such as rapid
rotations~\citep{Chakrabarti:2014,Martinon:2014,Cipolletta:2015,Breu:2016,Koliogiannis:2020,Khadkikar:2021,Annala:2021},
differential
rotations~\citep{Bretz:2015,Wendh:2019,Zhaotq:2022,Kuan:2022}, 
finite temperatures~\citep{Martinon:2014,Marques:2017,Lenka:2018,Stone:2019,Riahi:2019,Raduta:2020,Largani:2022},
strong magnetic fields~\citep{Haskell:2013,Soldateschi:2021}, and
within alternative theories of
gravity~\citep{Staykov:2016,Doneva:2016,Danchev:2021} (for a review,
see Ref.~\citep{Yagi:2017}). The recent work includes the study of
universal relations between the members of the $I$-Love-$Q$ triple and
compactness $C$ (which refers to the equatorial compactness of the star) 
for slowly rotating 
CSs~\citep{Bejger:2002,Lattimer:2004,Urbanec:2013,Baubock:2013,Maselli:2013}, 
and their extension to rapidly rotating
stars~\citep{Chakrabarti:2014,Breu:2016}. A different class of
universalities can be established between the parameters of static and
rotating CSs. For example, the redshift and maximum
mass~\citep{Lattimer:2004,Riahi:2019,Annala:2021}. Binary neutron
stars merger simulations have also revealed some relations between the 
properties of the dynamical ejecta and the binary parameters, such as 
the mass ratio and tidal deformability~\citep{Nedora:2021}.

\begin{figure*}[tb]
\centering
\includegraphics[width = 0.98\textwidth]{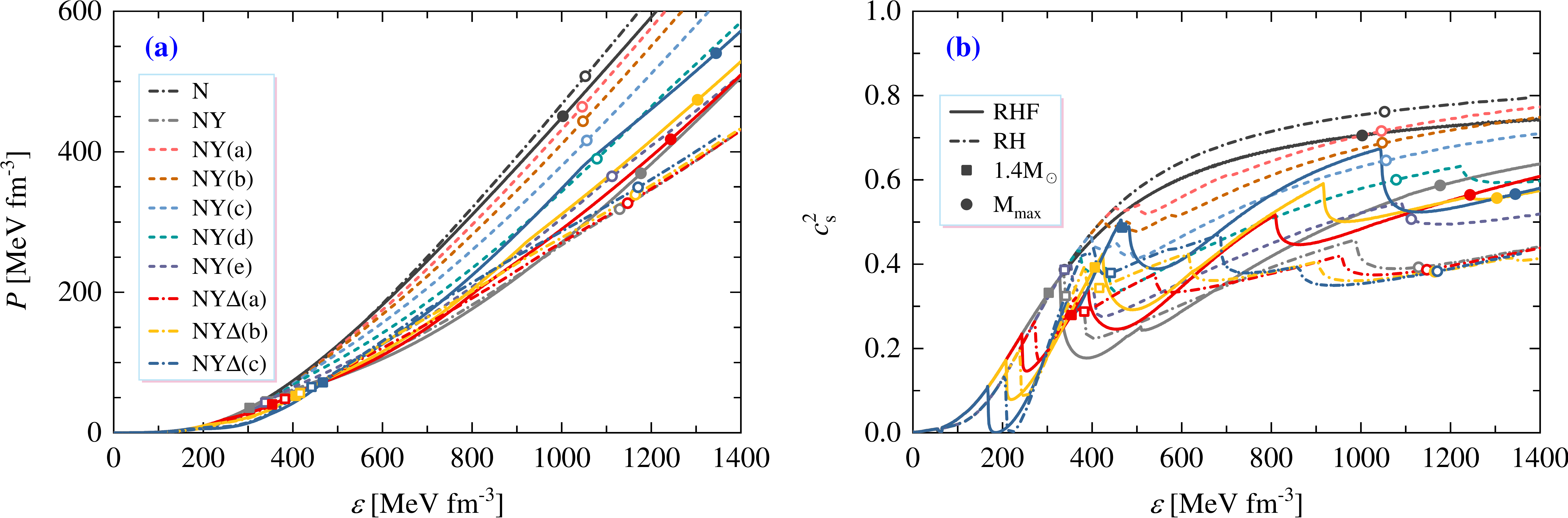}
\caption{EoSs for stellar matter featuring different compositions,
i.e., nucleonic (N), hyperonic (NY), and hyperon-$\Delta$ admixed 
(NY$\Delta$) one (panel a), and the associated speed of sound squared
$c^2_s$ (panel b). The results are obtained using both the RHF and RH 
approaches. The positions for canonical-mass and maximum-mass 
configurations are marked by squares and circles.}
\label{fig:EOS}
\end{figure*}

The purpose of this work is to test various universal relations
for static and rotating CSs using EoSs which account for the presence 
of heavy baryonic degrees of freedom in ultra-dense matter. To date, 
only a handful of studies have tested universal relations using such EoS. 
This was done for hyperonic matter in
Refs.~\citep{Yagi:2017,Marques:2017,Lenka:2018,Weijb:2018,Stone:2019,Khadkikar:2021}
and in the case of hyperonic $\Delta$-admixed matter in
Ref.~\citep{Raduta:2020}. For our purposes, we utilize the EoSs which
were previously derived within covariant density functional (CDF) theory
including hyperonic and $\Delta$-admixed matter at high
densities~\citep{Lijj:2018b,Lijj:2019a,Lijj:2019b,Sedrakian:2020prd,Sedrakian:2020epj,Sedrakian:2022epj}. 
Our EoSs are tuned to satisfy the astrophysical constraints, specifically 
the mass-radius ranges from NICER
inferences~\citep{NICER:2019a,NICER:2019b,NICER:2021a,NICER:2021b},
tidal deformability constraints from the GW event
GW170817~\citep{GW170817,LVC:2017}, and the symmetry energy constraints
from neutron skin thickness experiment
PREX-II~\citep{PREX:2021,Reed:2021,Reinhard:2021,Essick:2021}. 
The onset of heavy baryons significantly increases the complexity of
the EOS, as can be seen by examining the variability of the speed of 
sound $c_s$ across the star and comparing it with the  smooth behavior 
observed for purely nucleonic EoS models~\citep{Malfatti:2020,Lijj:2020}. 
It appears to be an important task to test available universal
relations in the case of matter with many particle thresholds and a
complex speed of sound behavior, given their significant practical
utility and importance.

Testing and validating the universal relations for EoS with 
heavy baryons expands the class of models of the EoS of matter that 
can be used in astrophysical scenarios which employ universal relations 
as an integral part of inference of properties of CSs. Such tests are 
also imperative in view of robust physical arguments in favor of the 
onset of heavy baryons in dense matter in CSs, as extensively discussed 
in the literature (for recent arguments see 
Refs.~\cite{Burgio:2021,Lattimer:2021,Sedrakian:2022ppnp} and references 
therein). Furthermore, the softening associated with heavy baryons is 
strongly constrained by the observations of two-solar-mass pulsars 
(addressed in the context of the hyperon puzzle) which allows us to 
have tight control over the uncertainties in the medium properties of 
hyperons in nuclear matter. Note that limiting the set of EoSs to those 
which satisfy the two-solar-mass constraint increases the accuracy to 
which the universal relations hold~\cite{Breu:2016}.

The paper is organized as follows. In Sec.~\ref{sec:EoS} we introduce 
the EoS model collection that we employed in our analyses. 
In Secs.~\ref{sec:Results-1} and~\ref{sec:Results-2}, we present the 
universal relations for isolated stars and the universal relation between 
the parameters of static and rapidly rotating stars, respectively. 
Finally, a summary of our results is provided in Sec.~\ref{sec:Conclusions}. 
Unless otherwise noted, we use geometric units (where $G=c=1$) throughout
this paper.

%-------------------------------------------------------------
\section{EoS for hadronic matter}
\label{sec:EoS}
%-------------------------------------------------------------
In the present analysis, we adopt the relativistic Hartree (RH) 
and Hartree-Fock (RHF)~\citep{Lijj:2018a} descriptions for dense stellar
matter, and consider three types of matter compositions:

\begin{figure*}[tb]
\centering
\includegraphics[width = 0.98\textwidth]{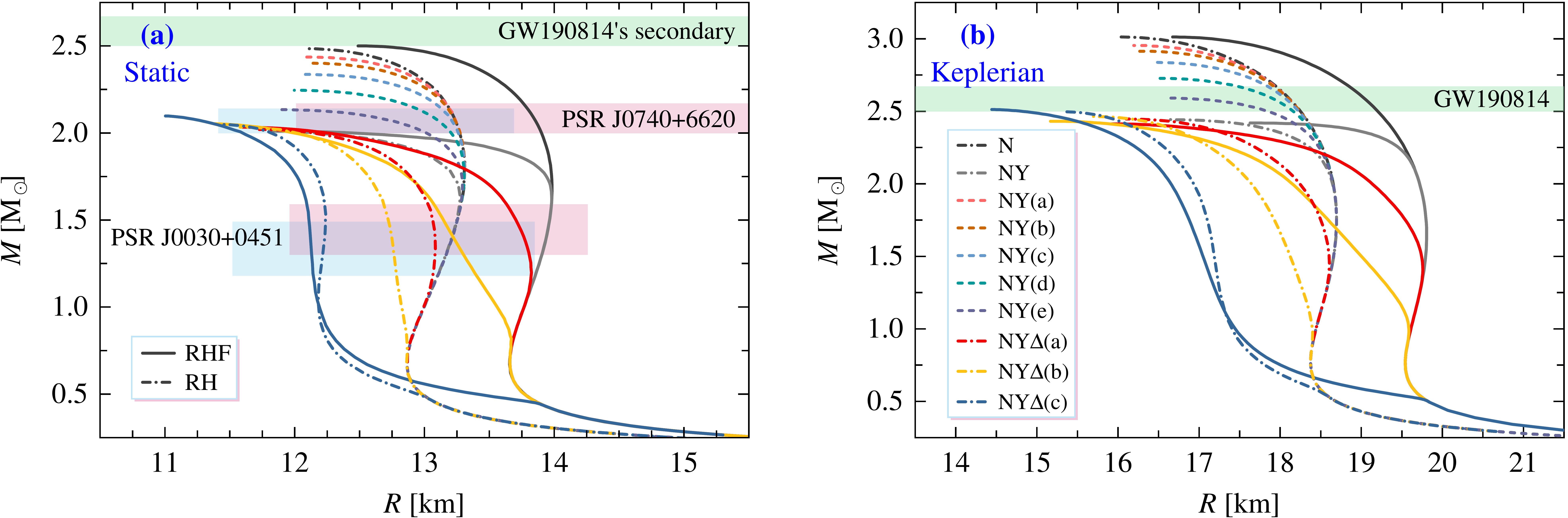}
\caption{Mass-radius relations of CSs in the static and maximally 
rotating (Keplerian) limits for various EoS models. The masses 
and radii for PSR J0030+0451~\citep{NICER:2019a,NICER:2019b} 
and PSR J0740+6620~\citep{NICER:2021a,NICER:2021b} ($68.3\%$ 
credible interval) are inferred from NICER data, and the mass 
range extracted for the secondary of the GW190814 
event~\citep{GW190814} is shown as well.
}
\label{fig:MR_relation}
\end{figure*}

1. Purely nucleonic EoS models. We use two representative
parametrizations of the nucleonic CDFs, specifically the RHF 
PKO3~\citep{Long:2008} and the RH DDME2~\citep{Lalazissis:2005} 
parametrizations.  Both of these parametrizations are accurately 
calibrated by the data on finite nuclei. The predicted maximum 
mass $M_{\rm max}$ and radius $R_{\rm 1.4}$ of canonical-mass 
$1.4\,M_{\odot}$ stars are
$M_{\rm max} = 2.49\,M_{\odot}$, $R_{\rm 1.4} = 13.96$~km for
PKO3~\cite{Lijj:2018a} and $M_{\rm max} = 2.48\,M_{\odot}$,
$R_{\rm 1.4} = 13.22$~km for DDME2~\citep{Lijj:2018a}. This class of
models is labeled as ``N". These nucleonic models produce massive 
neutron stars, which guarantees that in the case of moderate softening 
due to the onset of hyperons, the maximum masses of hyperonic stars
reach a value of $2\,M_{\odot}$ as required by the observations. 

2. Hyperonic EoS models. We extend the two nucleonic models mentioned 
above to the hyperonic sector by adjusting the meson-hyperon coupling 
constants to reproduce the empirical potentials of hyperons in nuclear 
matter. As a result, we are able to generate hyperonic stars with masses 
around $2\,M_\odot$. In particular, for the RH model, the vector 
meson-hyperon couplings are fixed by the SU(6) spin-flavor symmetric model,
with the scalar $\sigma$-meson-hyperon couplings defined by their ratios 
$R_{\sigma\Lambda}=0.6105$, $R_{\sigma\Sigma}=0.4426$, and 
$R_{\sigma\Xi}=0.3024$ to the corresponding nucleonic couplings 
(see Ref.~\citep{Sedrakian:2022ppnp} for a discussion). For the RHF 
model, the vector meson-hyperon couplings are determined by the SU(3) 
flavor symmetric model, to obtain more repulsion in the hyperonic sector 
to counterbalance the softening due to the Fock terms~\citep{Lijj:2018a}.
These two hyperonic EoS models are labeled as ``NY". Further EoS models 
that will be used below are based on the RH model but have broken SU(6)
quark symmetry in the vector meson sector, as explained in
Ref.~\citep{Lijj:2022}. This allows us to have hyperonic models with 
maximum masses in the range
$2.1 \lesssim M_{\rm max} \lesssim 2.4\,M_{\odot}$. This class of
models is labeled as ``NY(a)-(e)".

3. Hypernuclear models with an $\Delta$-admixture. These models are
the same as the two hyperonic models ``NY" above but include in addition the
quartet of spin-3/2 $\Delta$ resonances. As no consensus has been
reached yet on the magnitude of the $\Delta$ potential in nuclear
matter, we take three suggested values for the depth of the $\Delta$ 
potential, $V_{\Delta}(\rho_{\rm sat}) = (1\pm 1/3)V_{N}(\rho_{\rm sat})$, 
where $V_{N}(\rho_{\rm sat})$ is the nucleonic potential at saturation
density. This class of models is labeled as ``NY$\Delta$(a)-(c)". 
The main difference caused by the inclusion of $\Delta$'s is the
reduction of the star's radius by up to 1-2~km at central stellar 
densities slightly above the nuclear saturation density~\cite{Lijj:2018b}.

Figure~\ref{fig:EOS}~(a) shows the EoSs included in our collection. 
The squared speed of sound, $c^2_s$, obtained with these models is shown 
in Fig.~\ref{fig:EOS}~(b). It is seen that EoSs containing only nucleonic
degrees of freedom have a $c^2_s$ that monotonically increase with 
increasing baryon number density $\rho$ (or energy density $\varepsilon$). 
The appearances of hyperons and $\Delta$ particles change the shape of 
the curves to non-monotonic forms which reflects the nucleation of new 
degrees of freedom. The onset of heavy baryons reduces the speed of sound 
abruptly. Interestingly, in the case of an early appearance of $\Delta$ 
particles (at $\rho \sim 1.5\rho_{\rm sat}$), the speed of sound drops 
to zero indicating a possible region of instability associated with 
a liquid-gas type phase transition~\citep{Raduta:2021}. 

Figure~\ref{fig:MR_relation} shows the mass-radius relations of static
and maximally rotating, hereafter referred also as {\it Keplerian},
CSs. Additionally, the results include (a) the 68.3\% credible interval
for the mass and radius estimates of PSR
J0030+0451~\citep{NICER:2019a,NICER:2019b} and PSR
J0740+6620~\citep{NICER:2021a,NICER:2021b} as well as (b) the range of
masses extracted for the secondary object in the GW190814
event~\citep{GW190814}. It is seen that the masses and radii of
static configurations based on our EoS collection cover ranges of
$2.0 \lesssim M_{\rm max} \lesssim 2.5\,M_{\odot}$ and
$12 \lesssim R_{1.4} \lesssim 14$~km, i.e., our EoSs predict models
that are consistent with the current astrophysical constraints.

Maximally rotating stars have masses approximately 20\% larger than their
static counterparts because the centrifugal force provides additional
support against the gravitational pull toward the center of the star.
Their equatorial radii are about 40\% larger than the radii of static 
stars. It is seen that the maximum values for mass and radius of 
Keplerian configurations in our model collection cover the ranges
$2.5 \lesssim M_{\rm max} \lesssim 3.0\,M_{\odot}$ and
$17 \lesssim R_{1.4} \lesssim 20$~km.

%-------------------------------------------------------------
\section{Universal relations for isolated CSs}
\label{sec:Results-1}
%-------------------------------------------------------------
In this section, we investigate whether the universal relations among
integral quantities such as the mass, radius, moment of inertia, and
quadrupole moment maintain their validity for static and rotating CSs 
with heavy baryons in their centers, in particular, those possessing 
hyperon-$\Delta$ admixed cores. 

\begin{figure}[b]
\centering
\includegraphics[width = 0.46\textwidth]{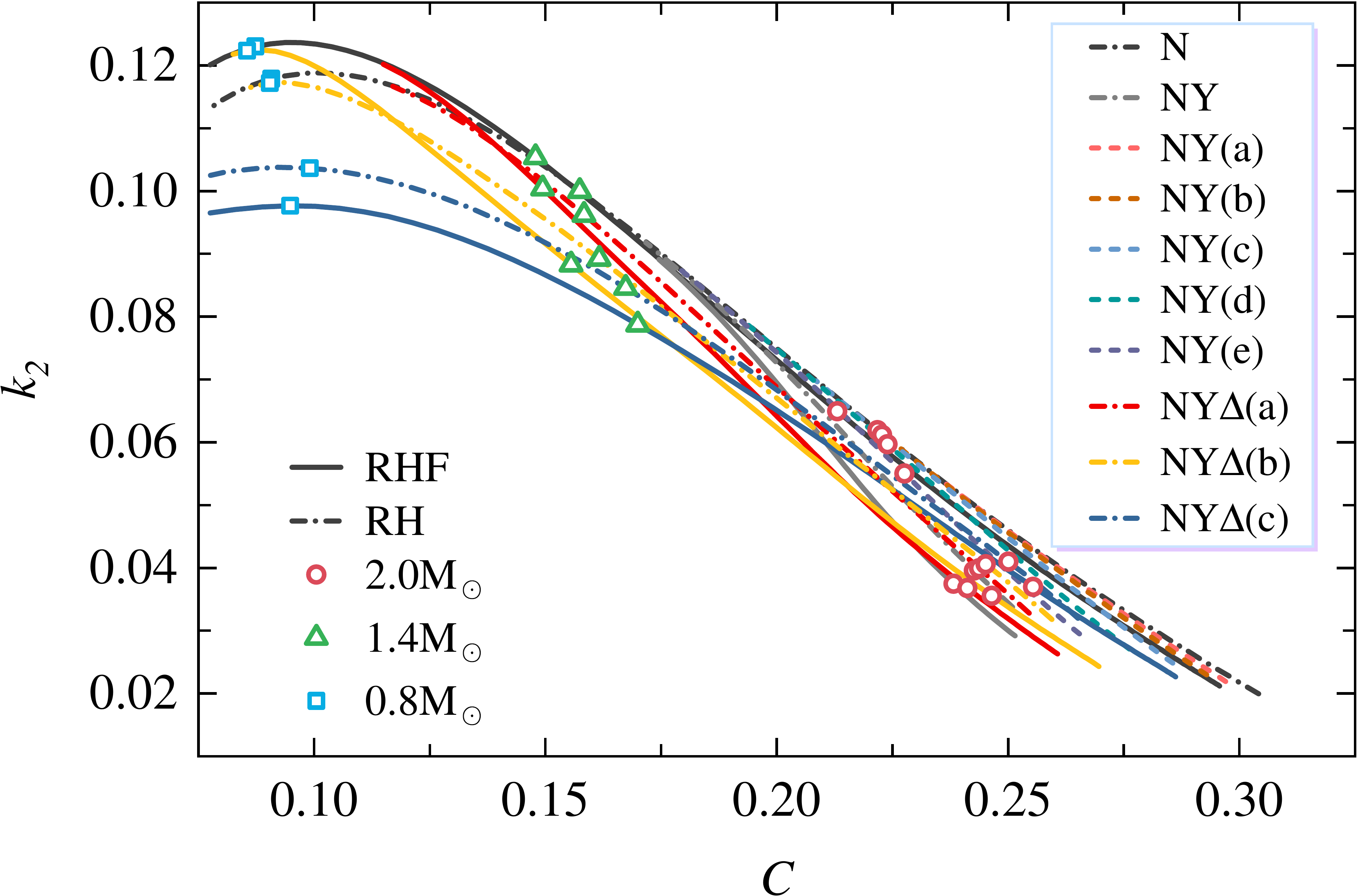}
\caption{The dimensionless tidal Love number $k_2$ 
as a function of compactness $C$ for hadronic EoS. 
The squares, triangles, and circles indicate configurations 
with $M =0.8$, 1.4 and $2.0\,M_{\odot}$, respectively.}
\label{fig:CKtwo_relation}
\end{figure}
\begin{table*}[tb]	
\centering
\caption{Estimated coefficients for the universal $C$-Love relation. 
The corresponding reduced chi-squared ($\chi^2_{\rm red}$) values 
are shown in the last column.}
\setlength{\tabcolsep}{9.4pt}
\begin{tabular}{ccccccccc}		
\hline\hline
$y$ & $x$ & $a_0$ & $a_1$ & $a_2$ & $a_3$ & $a_4$ & $\chi^2_{\rm red}$ \\
\hline
\multirow{2}*{$C$} & \multirow{2}*{$\Lambda$} &
                     $3.63278\times 10^{-1}$ & $-3.84806\times 10^{-2}$ & 
                     $1.77351\times 10^{-3}$ & $-1.78329\times 10^{-4}$ &         
                     $1.03106\times 10^{-5}$ & $ 7.82102\times 10^{-6}$ \\
        &          & $3.63958\times 10^{-1}$ & $-3.74047\times 10^{-2}$ &
                     $8.87636\times 10^{-4}$ & ~                        &
                     ~                       & $ 8.22215\times 10^{-6}$ \\ 
\hline\hline            
\end{tabular}
\label{table-CLove}
\end{table*}
%

%----------------------------------------------
\subsection{Static and slowly rotating CSs}
\label{sec:Slow}
\begin{figure}[b]
\centering
\includegraphics[width = 0.46\textwidth]{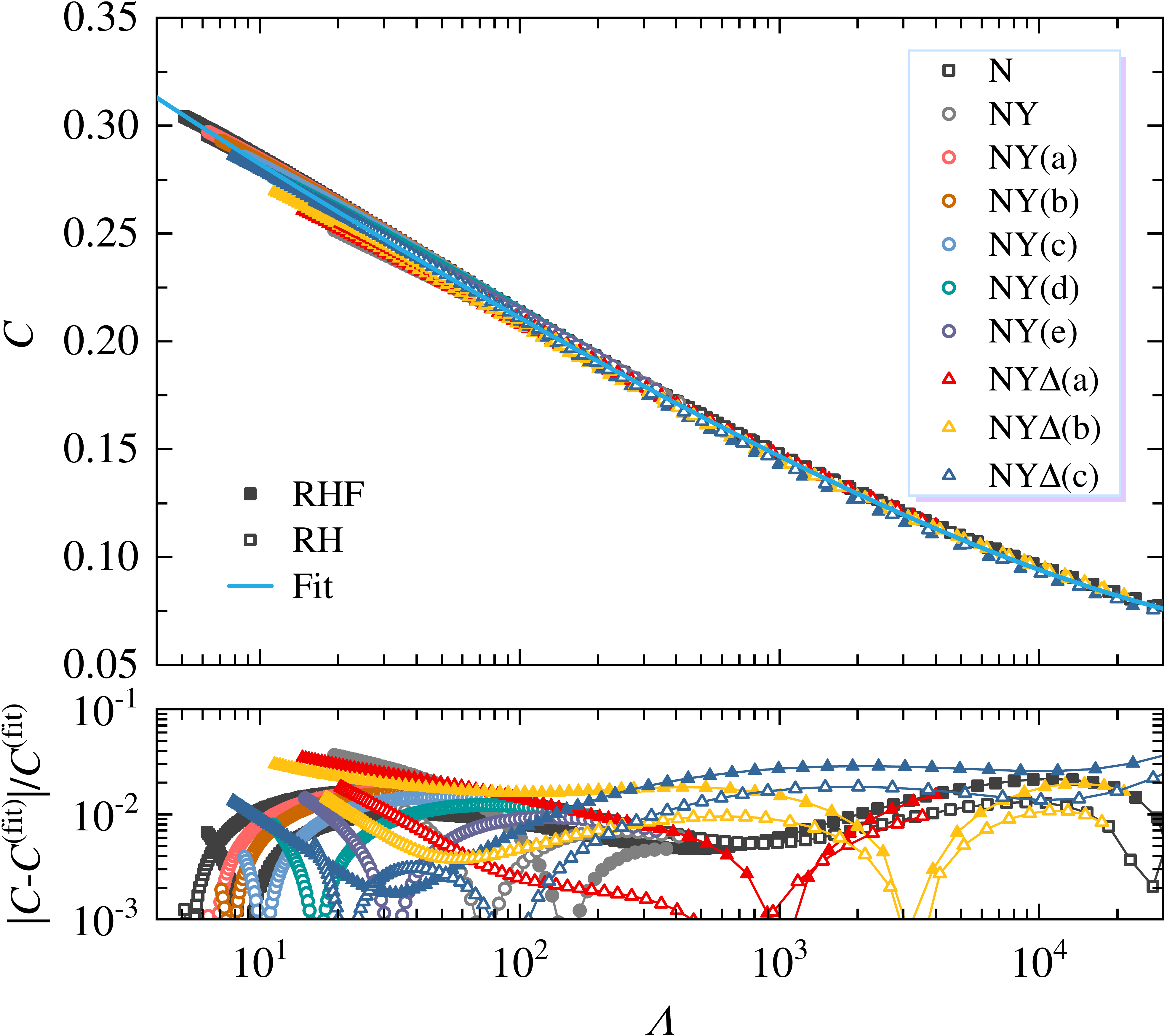}
\caption{The $C$-Love relations for static hadronic CSs.
(Top) Universal relations for various EoS models together with their 
fitting curves. (Bottom) Fractional errors between the fitting curve 
and numerical results.}
\label{fig:LamC_relation}
\end{figure}

%----------------------------------------------
The tidal deformability of static CS scales as $\Lambda\propto k_2C^{-5}$, 
according to Eq.~\eqref{eq:k2_equation}. The analytical form of the Love 
number $k_2$ is complicated, but the numerical evaluations of $k_2$ shown 
in Fig.~\ref{fig:CKtwo_relation} for $C \gtrsim 0.1$ (which is equivalent 
to $M \gtrsim 1.0\,M_{\odot}$) show that $k_2$ scales approximately as 
$C^{-1}$ and saturates for $C \approx 0.1$~\citep{Postnikov:2010,Lijj:2019a}.  
Therefore, the scaling becomes $\Lambda \propto C^{-6}$ or $C \propto \ln \Lambda$ 
for masses in the interval $1.0 \lesssim M \lesssim 2.0\,M_{\odot}$,
which is the mass range of phenomenological interest. The latter scaling 
behavior can be clearly observed in Fig.~\ref{fig:LamC_relation}, 
where the values of $\Lambda$ are shown on a logarithmic scale.

Building on the aforementioned scaling, one could hypothesize a scaling 
formula like
%-----------------------------------
\begin{align}\label{Eq:LC4_relation}
C = \sum^m_{n = 0} a_n \,(\ln \Lambda)^n,
\end{align}
%-----------------------------------
which was explored in Ref.~\citep{Maselli:2013} for $m=2$. The author 
of that paper used three EoS models (including a hybrid star model) with
a mass interval $1.2 \leqslant M \leqslant 2.0\,M_{\odot}$, and found 
the relation to be accurate up to 2\%. Later, this hypothesis was 
confirmed by Yagi et al. in Ref.~\citep{Yagi:2017} through the 
examination of a comprehensive collection of 25 hadronic EoS models,
which included 5 hyperonic models, and found that the maximum deviation 
is somewhat larger $\sim 6.5\%$. Recently, the same relation was tested 
for hot EoS appropriate for proto-neutron stars~\citep{Raduta:2020} and 
was found to hold with the same level of accuracy for stars with fixed 
entropy per baryon and lepton fraction, i.e., fixed thermodynamic conditions. 
Below we will utilize Eq.~\eqref{Eq:LC4_relation} truncated at $m = 4$, 
which is a polynomial degree employed in other universal relations examined 
in this work as well. 

Figure~\ref{fig:LamC_relation} shows $C$-$\ln \Lambda$ relation for
our collection of EoS in the case of static CSs according to
Eq.~\eqref{Eq:LC4_relation} with $m=4$. The bottom panel presents 
the fractional difference between the data and the fit. It is seen 
that the deviation is at the level of a few percent. The best-fit 
coefficients are summarized in Table~\ref{table-CLove} for
$m=2$ and $m=4$. The coefficients for $m=2$ obtained for our
collection of EoSs are in agreement with those found in
Ref.~\citep{Yagi:2017}.

\begin{figure*}[tb]
\centering
\includegraphics[width = 0.98\textwidth]{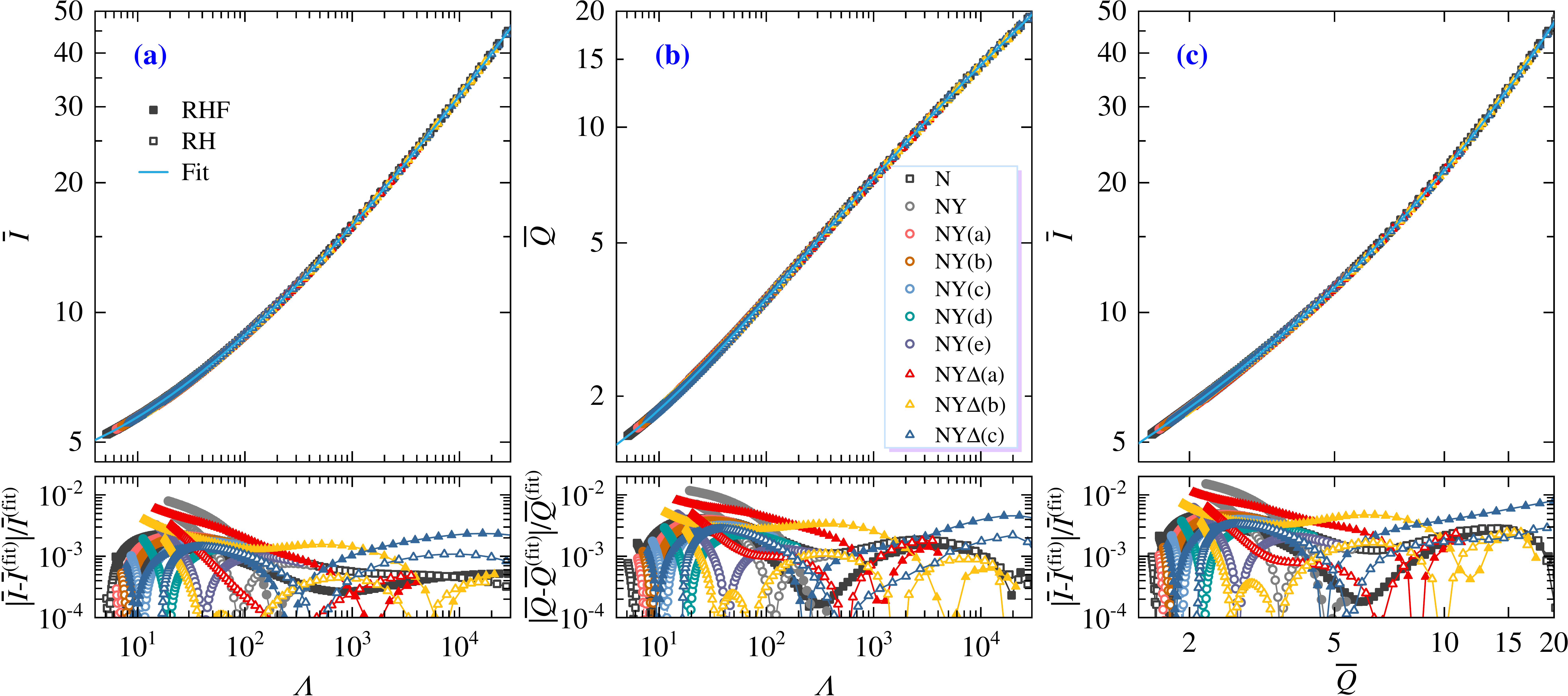}
\caption{The $\bar{I}$-Love-$\bar{Q}$ relations for hadronic CSs 
within the slow-rotation approximation. (Top) Universal relations 
for various EoS models together with their fitting curves. (Bottom) 
Fractional errors between the fitting curves and numerical results.}
\label{fig:ILQ_relation}
\end{figure*}
\begin{table*}[htb]	
\centering
\caption{
Estimated coefficients for the universal $\bar{I}$-Love, $\bar{Q}$-Love and 
$\bar{I}$-$\bar{Q}$ relations obtained in the slow-rotation approximation.
The corresponding reduced chi-squared ($\chi^2_{\rm red}$) values are shown 
in the last column.}
\setlength{\tabcolsep}{9.6pt}
\begin{tabular}{ccrrrrrc}		
\hline\hline
$y$ & $x$ & $a_0$ & $a_1$ & $a_2$ & $a_3$ & $a_4$ & $\chi^2_{\rm red}$ \\ 
\hline	                   
$\bar{I}$ & $\Lambda$ & $1.49834\times 10^{0}$  &$ 5.83188\times 10^{-2}$ &
                        $2.26613\times 10^{-2}$ &$-7.15384\times 10^{-4}$ &
                        $8.60779\times 10^{-6}$ &$ 2.91090\times 10^{-6}$ \\                   
$\bar{Q}$ & $\Lambda$ & $1.95655\times 10^{-1}$ &$ 9.02591\times 10^{-2}$ &
                        $4.87888\times 10^{-2}$ &$-4.39624\times 10^{-3}$ &
                        $1.30998\times 10^{-4}$ &$ 8.02989\times 10^{-6}$ \\
$\bar{I}$ & $\bar{Q}$ & $1.39803\times 10^{0}$  &$ 5.35419\times 10^{-1}$ &
                        $3.82137\times 10^{-2}$ &$ 1.81689\times 10^{-2}$ &
                        $1.92556\times 10^{-4}$ &$ 1.30642\times 10^{-5}$ \\   
\hline\hline       
\end{tabular}
\label{table-IloveQ}
\end{table*}

We now turn to the universal relations associated with 
the $I$-Love-$Q$ relation for our collection of EoSs.
Before showing the results, let us note that for a uniform 
Newtonian star one has~\citep{Yagi:2013b}
%--------------------------------------------
\begin{align}\label{EQ:Newtonian_limit_1}
\Lambda = \frac{1}{2} C^{-5}, \quad 
\bar{I} = \frac{2}{5} C^{-2}, \quad 
\bar{Q} = \frac{25}{8} C^{-1},
\end{align}
%--------------------------------------------
which translate into
%--------------------------------------------
\begin{align}\label{EQ:Newtonian_limit_2}
\bar{I} \propto \Lambda^{2/5}, \qquad 
\bar{Q} \propto \Lambda^{1/5}, \qquad 
\bar{Q} \propto \bar{I}^{1/2}.
\end{align}
%--------------------------------------------
Equations~\eqref{EQ:Newtonian_limit_1} 
and~\eqref{EQ:Newtonian_limit_2}
provide useful guidance for choosing 
the functional forms of universal relations 
for relativistic CSs.

The universal relations of $\Lambda$, $\bar{I}$, 
and $\bar{Q}$ can be explored using the scheme 
suggested in Refs.~\citep{Yagi:2013a,Yagi:2013b},
%--------------------------
\begin{align}\label{Eq:ILQ_relation}
\ln y  = \sum^4_{n = 0} a_n \, (\ln x)^n,
\end{align}
%--------------------------
where the pairs $(x, y)$ represent $(\Lambda,\,\bar{I})$,
$(\Lambda,\,\bar{Q})$ and $(\bar{Q},\,\bar{I})$.
Figure~\ref{fig:ILQ_relation} shows these three combinations 
for CSs containing heavy baryons together with the fits according 
to Eq.~\eqref{Eq:ILQ_relation}. The bottom panels of this figure 
show the fractional differences between the data and the fits. 
The moment of inertia and the quadrupole moment were computed 
assuming the slow rotation approximation up to second order. 
It is seen that the absolute fractional differences of these 
relations are $\lesssim 1\%$ for all three relations. The best-fit 
coefficients with Eq.~\eqref{Eq:ILQ_relation} are summarized in 
Table~\ref{table-IloveQ}, which are consistent with those found 
in Ref.~\citep{Yagi:2017}.

\begin{figure*}[tb]
\centering
\includegraphics[width = 0.98\textwidth]{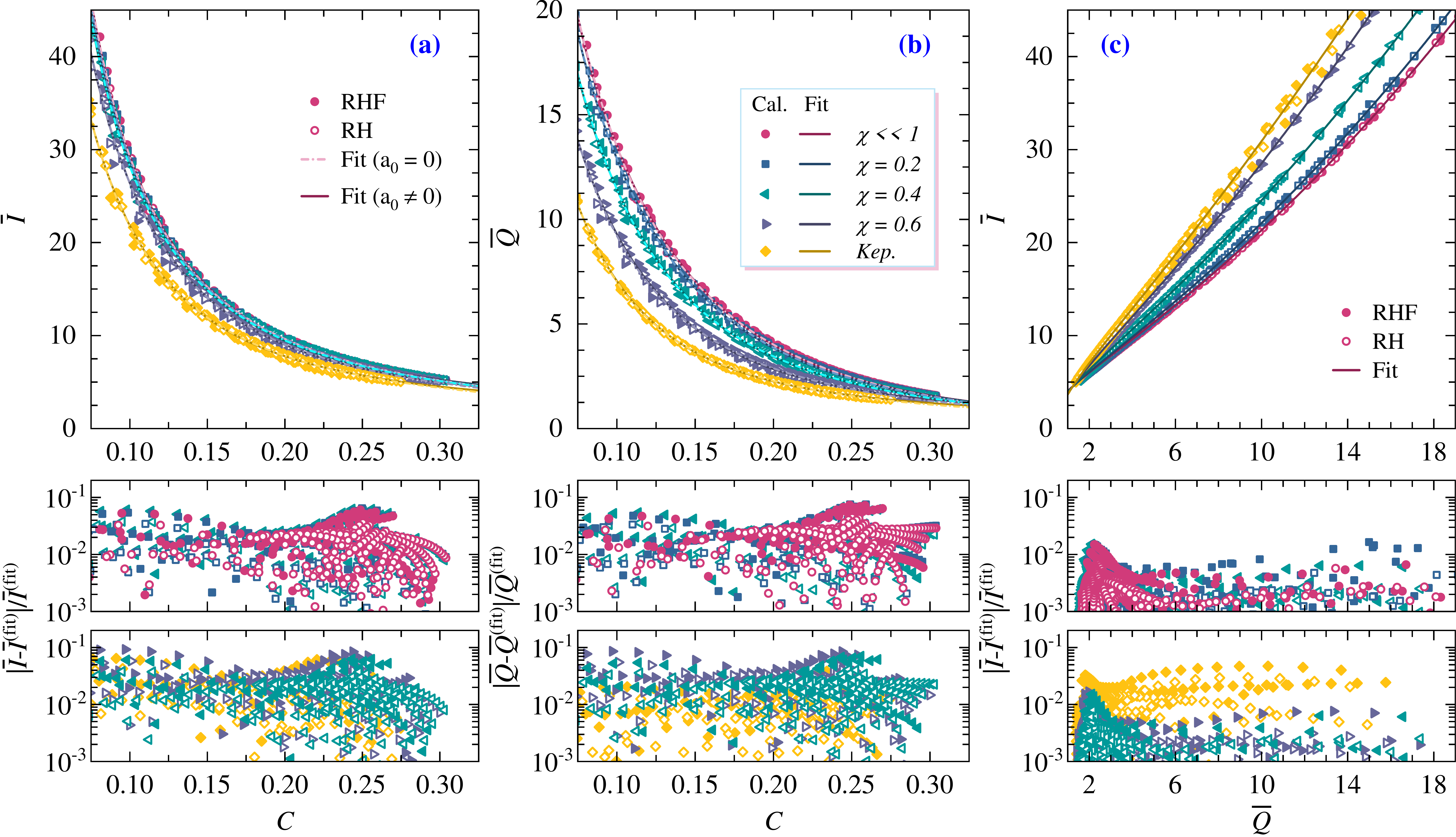}
\caption{The $\bar{I}$-$C$-$\bar{Q}$ relations for rotating 
hadronic CSs along sequences with constant spin parameter $\chi$.
(Top) Universal relations for various EoS models together with 
the fitting curves. (Bottom) Fractional errors between the 
fitting curves and numerical results.}
\label{fig:ICQ_rotation}
\end{figure*}

Because of the universality of the $C$-$\ln \Lambda$ relation 
and the universality among $\bar{I}$-$\Lambda$-$\bar{Q}$, it can 
be postulated that the relations $\bar I$-$C$ and $\bar Q$-$C$ 
also exhibit universality. The $\bar{I}$-$C$ and $\bar{Q}$-$C$ 
universal relations were investigated in 
Refs.~\citep{Yagi:2013b,Urbanec:2013,Breu:2016,Raduta:2020,Khadkikar:2021}
by exploring their functional forms with the inverse compactness 
$C^{-1}$ serving as the independent variable. Below, we will 
concentrate on this dependence, i.e.,
% ----------------------------------
\begin{align}\label{Eq:ICQ_relation}
\bar{I} = \sum^{m}_{n = 0} a_n \,(C^{-1})^n, \quad 
\bar{Q} = \sum^{m}_{n = 0} a_n \,(C^{-1})^n ,
\end{align}
with $m=4$.
%----------------------------------
For a large set of EoS the relation for $\bar I$ was found to be 
universal, with relative deviations being on the order of 10\%. 
Reference~\citep{Raduta:2020} considered also the universality
of the second relation in Eq.~\eqref{Eq:ICQ_relation} for $m=3$.
These universal relations~\eqref{Eq:ICQ_relation} where extended to
a set of finite-temperature hadronic EoS models (including two 
hyperonic and one hyperon-$\Delta$ admixed EoS models based on the 
DDME2 parametrization) in the case of fixed entropy per baryon 
and fixed lepton fraction. The universality holds with the same 
level of accuracy for both cold and hot stars, provided the 
thermodynamic parameters within the EoS are the same. 

\begin{table*}[!]	
\centering
\caption{
Estimated coefficients for the universal $\bar I$-$C$, $\bar Q$-$C$, 
and $\bar I$-$\bar Q$ relations. The corresponding reduced chi-squared 
($\chi^2_{\rm red}$) values are shown in the last column.}
\setlength{\tabcolsep}{6.8pt}
\begin{tabular}{ccccrrrrc}		
\hline\hline      
$y$ &$x$ & $\chi$ & $a_0$ & $a_1$ & $a_2$ & $a_3$ & $a_4$ & $\chi^2_{\rm red}$ \\ 
\hline
$\bar{I}$ & $C$  & $\ll 1$ &$ 2.78328\times 10^{0}$  &$-9.28412\times 10^{-1}$ &
                            $ 6.01079\times 10^{-1}$ &$-3.26388\times 10^{-2}$ &
                            $ 8.30545\times 10^{-4}$ &$ 6.51045\times 10^{-2}$ \\
         ~&     ~&        ~&                        ~&$ 7.52174\times 10^{-1}$ &
                            $ 2.49719\times 10^{-1}$ &$-2.52963\times 10^{-3}$ &
                            $-7.00854\times 10^{-5}$ &$ 6.68638\times 10^{-2}$ \\               
         ~&     ~&   $0.2$ &$ 2.40576\times 10^{0}$  &$-6.98316\times 10^{-1}$ & 
                            $ 5.53941\times 10^{-1}$ &$-2.92220\times 10^{-2}$ &
                            $ 7.29404\times 10^{-4}$ &$ 1.01976\times 10^{-1}$ \\
         ~&     ~&        ~&                        ~&$ 7.15632\times 10^{-1}$ &
                            $ 2.66210\times 10^{-1}$ &$-5.15035\times 10^{-3}$ &
                            $ 2.40009\times 10^{-5}$ &$ 1.03497\times 10^{-1}$ \\
         ~&     ~&   $0.4$ &$ 2.28723\times 10^{0}$  &$-6.25782\times 10^{-1}$ & 
                            $ 5.35113\times 10^{-1}$ &$-2.81811\times 10^{-2}$ &
                            $ 6.89230\times 10^{-4}$ &$ 1.22620\times 10^{-1}$ \\
         ~&     ~&        ~&                        ~&$ 7.09692\times 10^{-1}$ &
                            $ 2.65274\times 10^{-1}$ &$-5.77664\times 10^{-3}$ &
                            $ 3.78995\times 10^{-5}$ &$ 1.23993\times 10^{-1}$ \\  
         ~&     ~&   $0.6$ &$ 2.03438\times 10^{0}$  &$-4.04730\times 10^{-1}$ & 
                            $ 4.57441\times 10^{-1}$ &$-2.16860\times 10^{-2}$ &
                            $ 4.36667\times 10^{-4}$ &$ 2.75660\times 10^{-1}$ \\
         ~&     ~&        ~&                        ~&$ 7.52295\times 10^{-1}$ & 
                            $ 2.30435\times 10^{-1}$ &$-3.43435\times 10^{-3}$ &
                            $-7.60088\times 10^{-5}$ &$ 2.76760\times 10^{-1}$ \\                         
         ~&     ~&    Kep. &$ 1.59880\times 10^{0}$  &$-6.12213\times 10^{-2}$ & 
                            $ 3.26296\times 10^{-1}$ &$-1.47631\times 10^{-2}$ &
                            $ 2.97495\times 10^{-4}$ &$ 1.43312\times 10^{-1}$ \\
         ~&     ~&        ~&                        ~&$ 7.53968\times 10^{-1}$ &
                            $ 1.83664\times 10^{-1}$ &$-4.57602\times 10^{-3}$ &
                            $ 4.39646\times 10^{-5}$ &$ 1.44235\times 10^{-1}$ \\  
\hline                 
$\bar{Q}$ &$C$   & $\ll 1$ &$ 1.09156\times 10^{0}$  &$-1.19359\times 10^{0}$  &
                            $ 5.12662\times 10^{-1}$ &$-3.61313\times 10^{-2}$ &
                            $ 9.18768\times 10^{-4}$ &$ 1.01870\times 10^{-2}$ \\
         ~&     ~&        ~&                        ~&$-5.34487\times 10^{-1}$ &
                            $ 3.74863\times 10^{-1}$ &$-2.43229\times 10^{-2}$ &
                            $ 5.65554\times 10^{-4}$ &$ 1.04575\times 10^{-2}$ \\                           
         ~&     ~&   $0.2$ &$ 0.83661\times 10^{0}$  &$-9.49862\times 10^{-1}$ & 
                            $ 4.42625\times 10^{-1}$ &$-3.01219\times 10^{-2}$ &
                            $ 7.41153\times 10^{-4}$ &$ 1.43124\times 10^{-2}$ \\
         ~&     ~&        ~&                        ~&$-4.58158\times 10^{-1}$ &
                            $ 3.42566\times 10^{-1}$ &$-2.17509\times 10^{-2}$ &
                            $ 4.95847\times 10^{-4}$ &$ 1.44944\times 10^{-2}$ \\  
         ~&     ~&   $0.4$ &$ 1.17854\times 10^{0}$  &$-9.62231\times 10^{-1}$ & 
                            $ 3.98228\times 10^{-1}$ &$-2.63341\times 10^{-2}$ &
                            $ 6.39743\times 10^{-4}$ &$ 1.43084\times 10^{-2}$ \\
         ~&     ~&        ~&                        ~&$-2.74102\times 10^{-1}$ &
                            $ 2.59188\times 10^{-1}$ &$-1.47897\times 10^{-2}$ &
                            $ 3.04132\times 10^{-4}$ &$ 1.46898\times 10^{-2}$ \\  
         ~&     ~&   $0.6$ &$ 1.38135\times 10^{0}$  &$-8.18944\times 10^{-1}$ & 
                            $ 3.03745\times 10^{-1}$ &$-1.83139\times 10^{-2}$ &
                            $ 4.03007\times 10^{-4}$ &$ 2.60640\times 10^{-2}$ \\
        ~&      ~&         ~&                       ~&$-3.33255\times 10^{-2}$ & 
                            $ 1.49608\times 10^{-1}$ &$-5.92101\times 10^{-3}$ &
                            $ 5.49006\times 10^{-5}$ &$ 2.66705\times 10^{-2}$ \\                         
         ~&     ~&    Kep. &$ 0.93212\times 10^{0}$  &$-3.89707\times 10^{-1}$ & 
                            $ 1.70288\times 10^{-1}$ &$-8.73106\times 10^{-3}$ &
                            $ 1.71199\times 10^{-4}$ &$ 1.51724\times 10^{-3}$ \\
         ~&     ~&        ~&                        ~&$ 8.55566\times 10^{-2}$ &
                            $ 8.71318\times 10^{-2}$ &$-2.79190\times 10^{-3}$ &
                            $ 2.33887\times 10^{-5}$ &$ 1.88169\times 10^{-3}$ \\      
\hline  
$\bar{I}$&$\bar{Q}$&$\ll 1$&$ 1.39803\times 10^{0}$  &$ 5.35419\times 10^{-1}$ &
                            $ 3.82137\times 10^{-2}$ &$ 1.81689\times 10^{-2}$ &
                            $ 1.92556\times 10^{-4}$ &$ 1.30642\times 10^{-5}$ \\
        ~&         ~& $0.2$&$ 1.38034\times 10^{0}$  &$ 5.85404\times 10^{-1}$ &
                            $ 1.91033\times 10^{-2}$ &$ 2.26290\times 10^{-2}$ &
                            $-3.53397\times 10^{-4}$ &$ 1.42634\times 10^{-5}$ \\  
        ~&         ~& $0.4$&$ 1.35669\times 10^{0}$  &$ 7.04071\times 10^{-1}$ &
                            $-2.36864\times 10^{-2}$ &$ 3.28473\times 10^{-2}$ &
                            $-1.86181\times 10^{-3}$ &$ 1.19826\times 10^{-5}$ \\  
        ~&         ~& $0.6$&$ 1.31300\times 10^{0}$  &$ 9.21151\times 10^{-1}$ &
                            $-1.28186\times 10^{-1}$ &$ 6.18742\times 10^{-2}$ &
                            $-5.65783\times 10^{-3}$ &$ 1.50910\times 10^{-5}$ \\  
        ~&         ~& Kep. &$ 1.29830\times 10^{0}$  &$ 1.02524\times 10^{0}$  &           
                            $-1.75735\times 10^{-1}$ &$ 7.80475\times 10^{-2}$ &
                            $-8.85528\times 10^{-3}$ &$ 1.78771\times 10^{-4}$ \\
\hline\hline
\end{tabular}
\label{table-ICQ}
\end{table*}

Figure~\ref{fig:ICQ_rotation} shows $\bar{I}$ and $\bar{Q}$ as a
function of $C$, and the $\bar{I}$-$\bar{Q}$ relation for sequences
computed using the slow-rotation approximation ($\chi \ll 1$),
together with rapidly rotating sequences with fixed spin parameters
$\chi$ ($\chi \equiv J/M^2$ with $J$ being the angular momentum of the
star), and the maximally rotating sequence. The subsequent subsections 
will discuss the latter two categories of sequences. In panels~(a) and 
(b), we present for each $\bar{I}$-$C$ or $\bar{Q}$-$C$ relation two 
fitting curves, one using Eq.~\eqref{Eq:ICQ_relation} and one dropping
the 0-th order term. In panel (c), each relation is shown by a single
curve given by Eq.~\eqref{Eq:ILQ_relation}. Table~\ref{table-ICQ} 
summarizes the coefficients for these universal relations, along with 
the corresponding reduced chi-squared $\chi^2_{\rm red}$ values (which 
are obtained by dividing the residual sum of squares by the degrees of 
freedom). The fractional differences are shown in the lower panels of 
Fig.~\ref{fig:ICQ_rotation} whereby those for panels~(a) and (b) are 
evaluated with respect to fits that use Eq.~\eqref{Eq:ICQ_relation}.
As can be seen in Fig.~\ref{fig:ICQ_rotation}, the maximum fractional
differences are close to 7-8\% for the $\bar{I}$-$C$ and $\bar{Q}$-$C$
relations, which can be compared to the $\bar{I}$-$\Lambda$ and
$\bar{Q}$-$\Lambda$ universal relations where the deviations were mostly
below~1\%.

%----------------------------------------------
\subsection{Rapidly rotating CSs}
\label{sec:Rapid}
%----------------------------------------------
%
\begin{figure}[b]
\centering
\includegraphics[width = 0.46\textwidth]{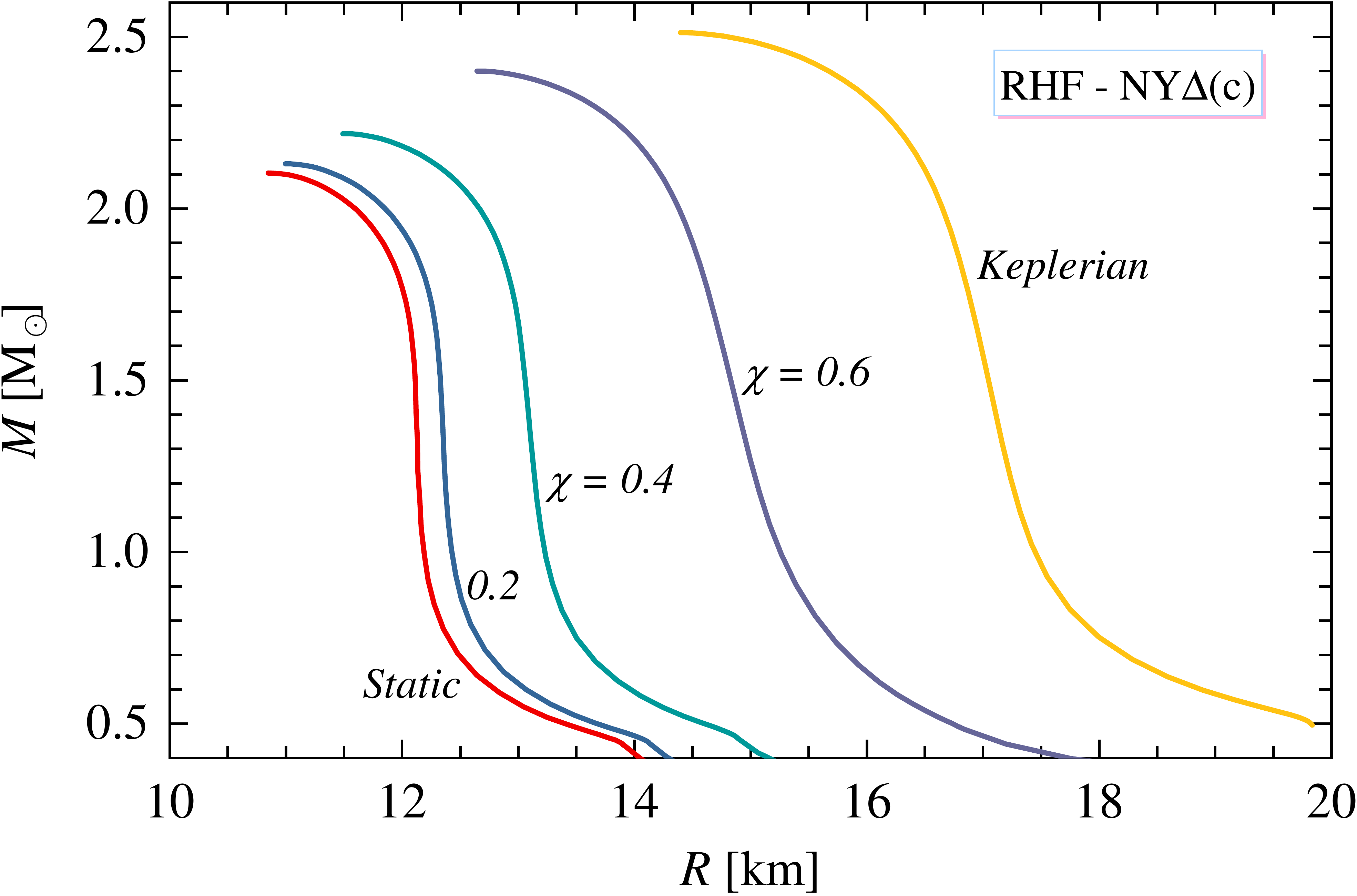}
\caption{The mass-radius relations of CSs for different rotation 
rates, namely static, rapidly rotating with constant spin 
parameters of $\chi = 0.2$, 0.4, and 0.6, as well as the maximum 
rotation rate (Keplerian limit), using the RHF EoS of NY$\Delta$(c).}
\label{fig:MR_rotation}
\end{figure}

We next test the validity of the universal relations among
$\bar{I}$, $\bar{Q}$, and $C$ for rapidly rotating CSs. Note
that compactness in this case is defined using the equatorial 
radius. In this case, the universal relations are commonly 
investigated for a sequence of stars with constant values of 
certain parameters characterizing the magnitude of rotation.

The universality of the $\bar{Q}$-$\bar{I}$ relation holds for 
stellar sequences with fixed dimensionless spin parameter
$\chi=J/M^2$~\citep{Pappas:2014} and $\tilde{f} = Rf$, where $f$ 
is the spin frequency~\citep{Chakrabarti:2014}, but not for stars 
with constant spin frequency $f$~\citep{Doneva:2013}. The
$\bar{Q}$-$\bar{I}$ relation was found for fixed $\chi$ and 
$\tilde f$ to be nearly independent of the EoS with a relative 
error $\sim 1\%$, i.e., with an error comparable to the slow-rotation 
case. In addition, Ref.~\citep{Breu:2016} found universality of 
$\bar{I}$-$C$ relation for rotating nucleonic CSs for three values 
of the spin parameter. The universal relations for hot and maximally 
fast rotating CSs were established in Ref.~\citep{Khadkikar:2021}.

Below we will show universal relations for sequences with constant
spin parameters of $\chi = 0.0,\, 0.2,\, 0.4$, and 0.6, as well as 
the Keplerian limit. The mass-radius relations of these sequences 
are shown in Fig.~\ref{fig:MR_rotation} for our RHF EoS model with
NY$\Delta$ composition. It is observed that these relations are
self-similar for different values of the spin parameter $\chi$.

Our results for $\bar{I}$-$C$-$\bar{Q}$ relations are presented in
Fig.~\ref{fig:ICQ_rotation} for the $\chi$ values quoted just above.  
As in the non-rotating case, the $\bar{I}$-$C$ and $\bar{Q}$-$C$ 
relations are fitted with two different functions: one is given by 
the full Eq.~\eqref{Eq:ICQ_relation} and the other with a vanishing 
zero-order term.  The fractional differences are estimated for the 
full expression. For the $\bar{Q}$-$\bar{I}$ relation the fitting 
curves are given by Eq.~\eqref{Eq:ILQ_relation}.

It is seen in Fig.~\ref{fig:ICQ_rotation} that the fractional
differences are comparable in magnitude for all values of the
spin parameter and for the static case. The $\bar{I}$-$C$ and
$\bar{Q}$-$\bar{I}$ relations for each sequence are self-similar 
and tend to a single value in the extreme ``black hole'' limit 
where $\bar{I} \to 4$ and $\bar{Q} \to 1$ when $C \to 0.5$. 
Again, as in the static case, the $\bar{Q}$-$\bar{I}$ relation 
holds to higher accuracy for all sequences. The maximum absolute 
fractional difference is between about 1-2\%, which is comparable 
to that for static stars.

\begin{figure}[b]
\centering
\includegraphics[width = 0.46\textwidth]{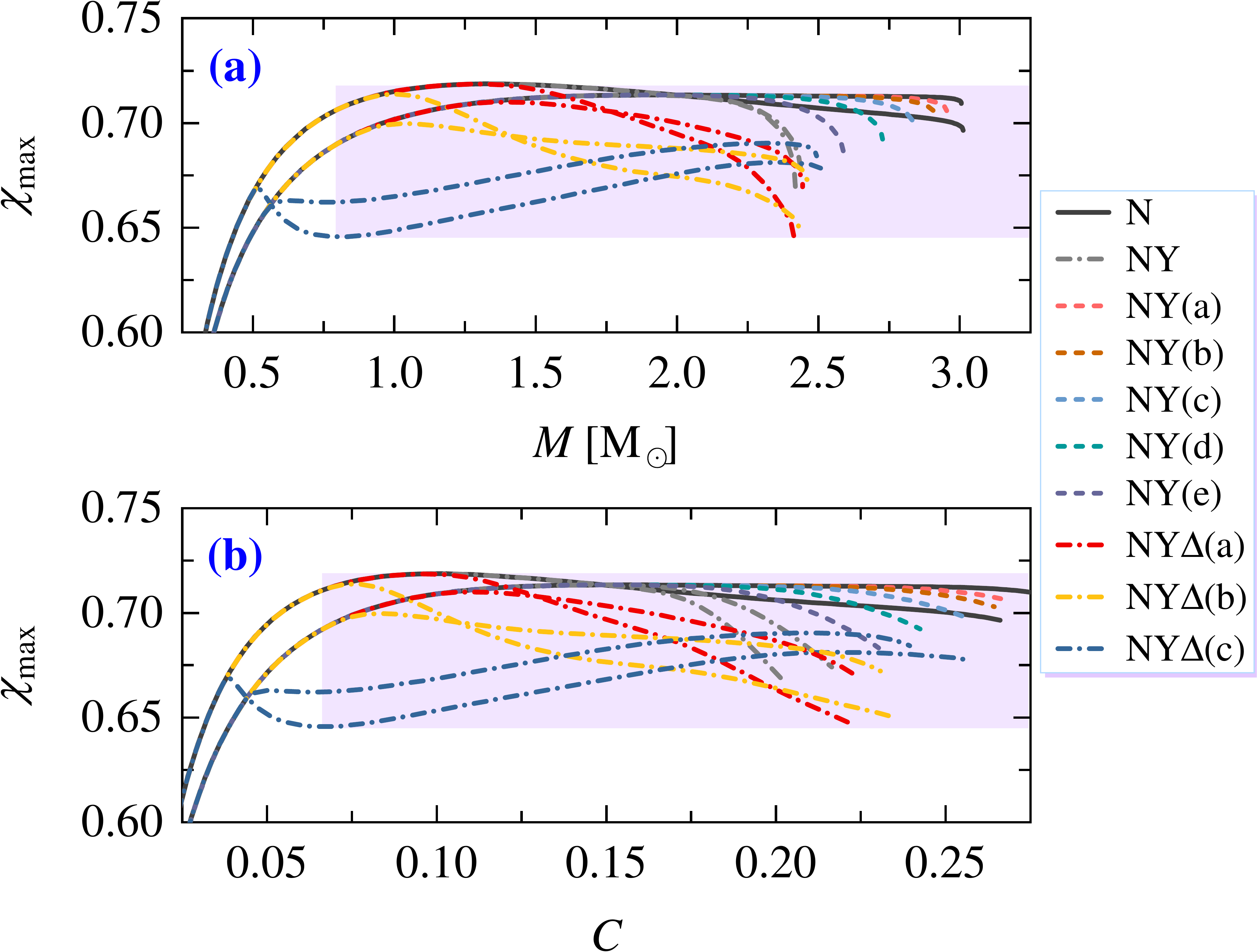}
\caption{The maximum spin parameter $\chi_{\rm max}$ as function 
of gravitational mass $M$ (panel a) and compactness $C$ (panel b)
for hadronic CSs.}
\label{fig:Jmax_Kep} 
\end{figure}

CSs reach the limit of stability for uniform rotation at the 
Keplerian limit, where the centrifugal and gravitational forces 
at the equator of the star are balanced. Beyond this frequency, 
mass is shed from the equator. A convenient parameter for locating 
the Keplerian limit is the spin parameter $\chi$~\citep{Lijj:2022,Lo:2011}. 
In Fig.~\ref{fig:Jmax_Kep} we plot the maximum spin parameter 
$\chi_{\rm max}$ versus gravitational mass and compactness for 
CSs constructed from our collection of EoS. It is seen that 
$\chi_{\rm max}$ increases with mass in the domain $M < 1.0\,M_{\odot}$ 
and then (a) stays constant for purely nucleonic EoS models; 
(b) it is reduced for models which contain hyperons; (c) the reduction 
is more pronounced for models that have in addition to hyperons a 
$\Delta$-resonance admixture in their cores. Thus, the $\chi_{\rm max}$ 
as a function of mass $M$ or compactness $C$ can no longer be monotonic; 
this is more prominent for the NY$\Delta$ models. The value of 
$\chi_{\rm max}(M)$ or $\chi_{\rm max}(C)$ is noteworthy in that 
all models fall within a narrow range of $0.64-0.72$, as indicated 
by the shaded region in Fig.~\ref{fig:Jmax_Kep}. The values of 
$\chi_{\rm max}$ derived for our EoS collection are consistent with
those found in Refs.~\citep{Cook:1994,Haensel:1995,Lo:2011,Lijj:2022}.

Figure~\ref{fig:ICQ_rotation} also shows $\bar{I}$-$C$-$\bar{Q}$
relations for Keplerian sequences.  In this case, the fractional
difference for the $\bar{I}$-$\bar{Q}$ relation increases to 5\%,
because $\chi_{\rm max} = 0.68 \pm 0.04$ for different EoS,
i.e., it is not constant anymore but lies in the indicated range. 
The fractional differences for the $\bar{I}$-$C$ and $\bar{Q}$-$C$ 
relations remain comparable to those for static and constant spin 
sequences. The fit coefficients in the universal relations for our 
sequences with constant $\chi$, including the Keplerain sequences, 
are summarized in Table~\ref{table-ICQ}. Also provided in this table 
are the corresponding reduced chi-squared ($\chi^2_{\rm red}$) values.

\begin{figure*}[tb]
\centering
\includegraphics[width = 0.98\textwidth]{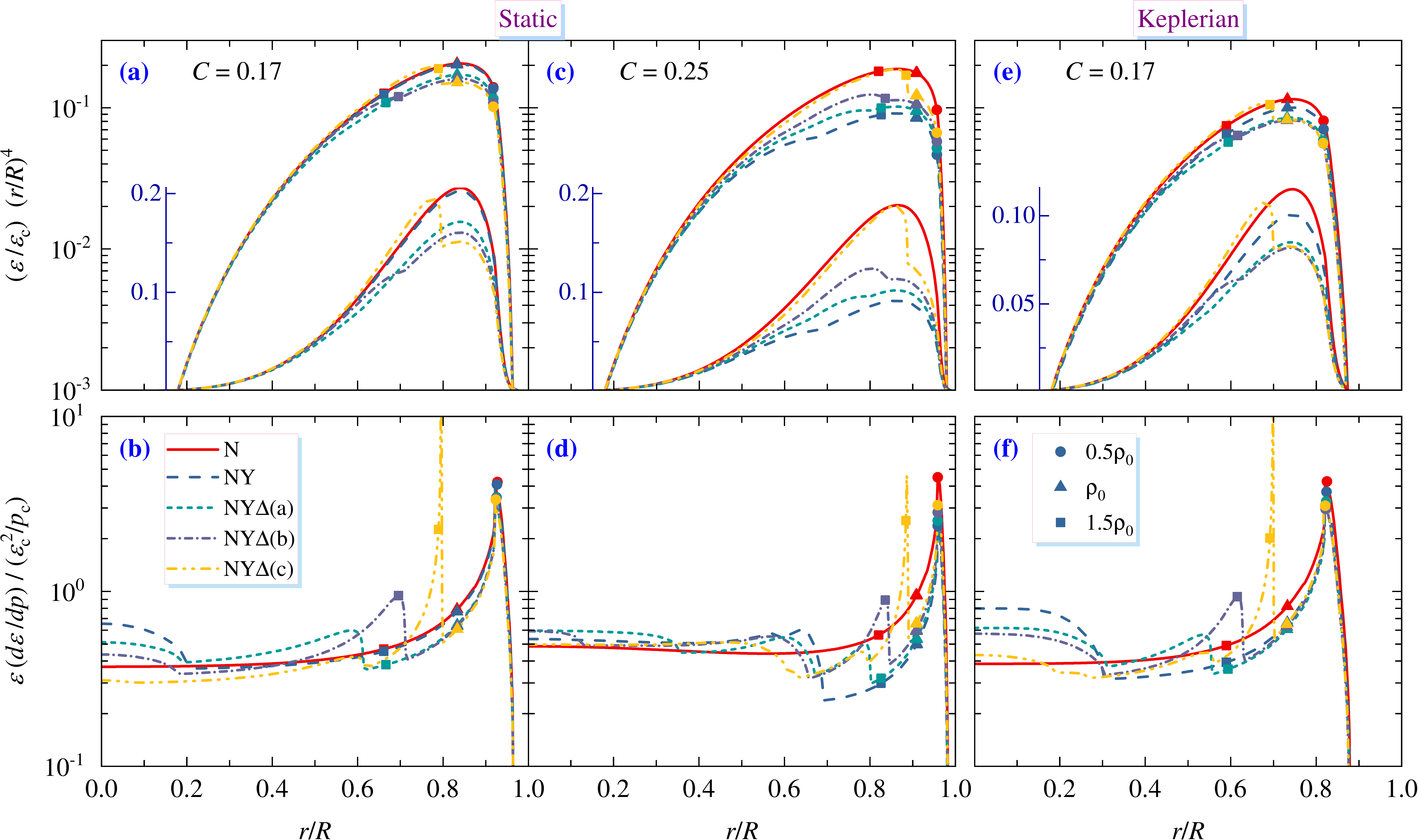}
\caption{The normalized radial profiles 
$[(\varepsilon/\varepsilon_c)/(r/R)^4$] and 
$[\varepsilon(d\varepsilon/dp)/(\varepsilon_{c}^{2}/p_c)]$ as 
a function of $r/R$ for static (left panels) CSs with compactness 
$C=0.17$, 0.25, and Keplerian rotating (right panels) CSs with $C=0.17$. 
The results are calculated with the RHF models for various particle 
compositions. The positions of the crust-core transition 
($\sim 0.5\,\rho_0$) and $1.0,\,1.5\,\rho_0$ are marked. 
In the top panels, the data are shown on a linear scale, too.  
}
\label{fig:Radial_profile-1}
\end{figure*}

The universalities shown in Fig.~\ref{fig:ICQ_rotation} allow 
(in principle) the determination of the radius of a pulsar, 
if the mass and one of the quantities $\bar{I}$ and $\bar{Q}$ 
are sufficiently well measured. The mass in binaries containing 
a neutron star (pulsar) can be measured by the relativistic 
Shapiro delay. Then, according to the definitions of $\bar{I}$ 
or $\bar{Q}$, the radius of the pulsar can be derived. We note, 
however, that the $\bar{I}$ and $\bar{Q}$ are largely degenerate 
for pulsars with a small value of spin or a large value of 
compactness. This makes it very challenging to measure the radius 
of a pulsar from the moment of inertia $I$, orquadrupole moment 
$Q$ independently.

%----------------------------------------------
\subsection{Radial profiles of CSs} 
%---------------------------------------------- 
Several hypotheses for the origin of universalities were put forward 
early on in Refs.~\citep{Yagi:2013a,Yagi:2013b}, but their physics 
remains a matter of debate. It has been pointed out that the contribution 
of the outer stellar layer, where the physics is mostly settled, may 
render the integral parameters insensitive to the details of the CS's 
physics at higher densities. It has also been pointed out that
universalities could be reminiscent of the no-hair theorems for black
holes. For models with compactness approaching the black-hole limit,
the details of the internal structure may become unimportant.
We will discuss these points below using our collection of EoSs.

\begin{figure*}[tb]
\centering
\includegraphics[width = 0.98\textwidth]{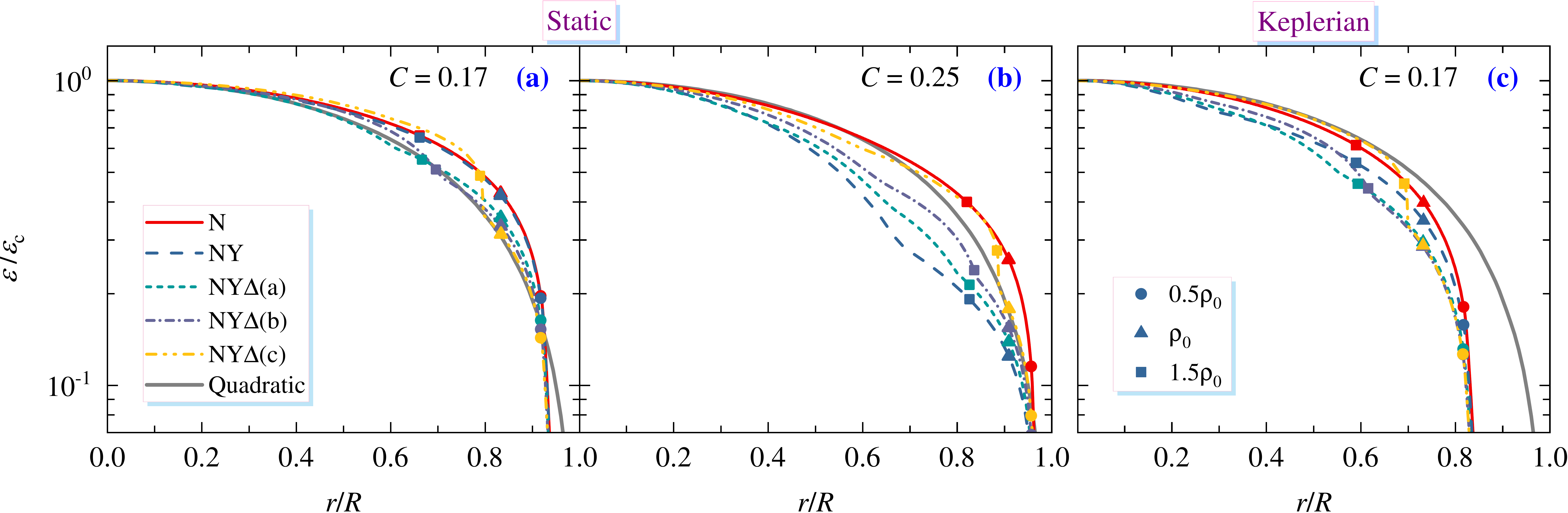}
\caption{The normalized radial density profiles
  $[\varepsilon/\varepsilon_c]$ as a function of $r/R$ for static 
  (left panels) CSs with compactness $C=0.17$, 0.25, and Keplerian
  rotating (right panels) stars with $C=0.17$. The profiles of a
  quadratic model for $\varepsilon/\varepsilon_c$ (see text) 
  are shown as well.}
\label{fig:Radial_profile-2}
\end{figure*}

To gain insight into the origin of universalities it is useful to
investigate the radial dependence of quantities of interest. In
Fig.~\ref{fig:Radial_profile-1}, we present the (normalized) radial
profiles $[\varepsilon r^4/\varepsilon_cR^4]$ and
$[\varepsilon(d\varepsilon/dp)/(\varepsilon_{c}^{2}/p_c)]$ for static
(left panels) CSs with compactnesses $C=0.17$, 0.25, as well as
Keplerian rotating (right panels) CSs with compactness $C=0.17$. Here
$\varepsilon_{c}$ and $p_{c}$ are the central energy density and the
corresponding pressure, respectively, and $R$ is the equatorial radius. 
These results were obtained using RHF EoS with various N, NY, and 
NY$\Delta$ compositions. The sudden changes in the slope of each the 
curve are associated with the crust-core transition and the onsets of
various species of heavy baryons. For static configurations, the
plotted quantity $[\varepsilon r^4]$ appears in the integrands of the 
moment of inertia and quadrupole moment in the Newtonian limit, whereas 
the quantity $[\varepsilon(d\varepsilon/dp) = \varepsilon c^2_s]$ 
appears in the integrand of tidal deformability~\cite{Yagi:2013b}.

The top panels of Fig.~\ref{fig:Radial_profile-1} show the quantity
$[\varepsilon r^4]$, which is relevant for $I$ and $Q$. It is seen in
Fig.~\ref{fig:Radial_profile-1}~(a), where $C=0.17$, that the profiles of
this quantity are the same for $r/R < 0.6$ and $r/R > 0.9$. Noticeable
differences appear within the range $0.6 \le r/R \le 0.9$ with the
maxima of the curves differing by up to 30\%. This range corresponds 
to the low-density regions around the nuclear saturation density, which 
dominantly contributes to $I$ and $Q$. For more compact stars with $C=0.25$, 
shown in Fig.~\ref{fig:Radial_profile-1}~(c), noticeable deviations 
appear already for $r/R \ge 0.5$, with variations of up to 50\% at the 
maxima located at $r/R \approx 0.8$. The profiles of Keplerian models 
with $C=0.17$, shown in panel (e), can now be compared to those of the 
static CSs with the same compactness. This comparison shows the same 
qualitative behavior but with lower values of the maxima.

The bottom panels of Fig.~\ref{fig:Radial_profile-1} show the profile
of $[\varepsilon(d\varepsilon/dp)]$ relevant for tidal deformability.
It is seen that for the case $C=0.17$, shown in 
Fig.~\ref{fig:Radial_profile-1}~(b), peaks appear close to the
crust-core transition and within the range
$0.6 \lesssim r/R \lesssim 0.8$ due to the onset of heavy baryons.
The radial profiles of more compact stars with $C=0.25$, shown in 
Fig.~\ref{fig:Radial_profile-1}~(d), have a similar structure to 
the previous case, with substantial (quantitatively similar) deviations 
between different EoSs in the range $0.6 \lesssim r/R \lesssim 0.9$.

Our results above provide no evidence that the $\bar I$-$C$-$\bar Q$
universality results from the independence of the profiles of the
integrands of these global quantities for our collection of EoSs in the
outer core and the crust. In fact, while the EoS in the low density
range $\rho/\rho_0 \le 1.5$ are similar, the profiles of relevant
quantities as a function of {\it normalized} radius, as illustrated in
Fig.~\ref{fig:Radial_profile-1}, are closely matched only in the
innermost core regions and the outermost crustal regions. Consider
now the interval $0.5 \lesssim r/R \lesssim 0.9$, where deviations are
observed. We note that the functional behavior (increase or decrease)
of $[\varepsilon r^4]$ with $r/R$ is similar to that of the quantity
$[\varepsilon(d\varepsilon/dp)]$.  This ``correlation'' is best seen
for $C=0.25$.  Therefore, one may conclude that the high level of
universality of $\bar I$-$\Lambda$-$\bar Q$ relations is mainly
attributed to the ``correlation'' between the underlying profiles of
$\bar{I}(\bar{Q})$-$C$ and $\bar\Lambda$-$C$, which leads to a cancellation
when combined.

Among the various suggestions proposed to explain the origin of
universality, Ref.~\citep{Yagi:2014} put forward the idea that the
assumption of self-similarity of isodensity contours in realistic
CSs, which can be approximated by elliptical contours, plays
a crucial role in the universality of $I$-Love-$Q$ relations.
Furthermore, in Ref.~\citep{Sham:2015} numerical evidence was presented
that demonstrated that the universality, which holds in the
incompressible limit and implies self-similar isodensity surfaces, 
is retained for modern realistic EoS. These EoS are known to be stiff
and, thus, have also self-similar isodensity surfaces. Another
hypothesis proposed in Ref.~\citep{Jiangnan:2020} suggests that the
universal relations arise from the fact that the energy density of
a realistic CS can be approximated as a quadratic function of the
normalized radius, given by
$\varepsilon (r/R) = \varepsilon_c \big[1-(r/R)^2\big]$, where
$\varepsilon_c$ is the central stellar energy density.

In Fig.~\ref{fig:Radial_profile-2}, we present the radial profiles 
of energy density $[\varepsilon/\varepsilon_c]$, together with the 
ones from the quadratic model. This allows one to account for the
difference between the realistic CS profile and that of the quadratic
model. It is seen that the density profiles of less compact static
stars, $C=0.17$, can be approximated with an accuracy of up to 30\% 
(except for the crust region) with the quadratic model, whereas for 
more compact static stars, e.g., for $C=0.25$, or for rapidly rotating 
stars, the profiles deviated strongly from a quadratic one.  

In closing, it is worth noting that the integrand of the mass $M$, 
which is given in terms of $[\varepsilon r^2]$, varies with $r/R$ 
like the normalized energy density shown in the top panels of 
Fig.~\ref{fig:Radial_profile-1}. Because mass enters the definitions 
of the dimensionless quantities $\bar{I}$ and $\bar{Q}$, the similarity 
in the variations of the integrands of these quantities may partially 
cancel each other out. 

\begin{figure}[b]
\centering
\includegraphics[width = 0.46\textwidth]{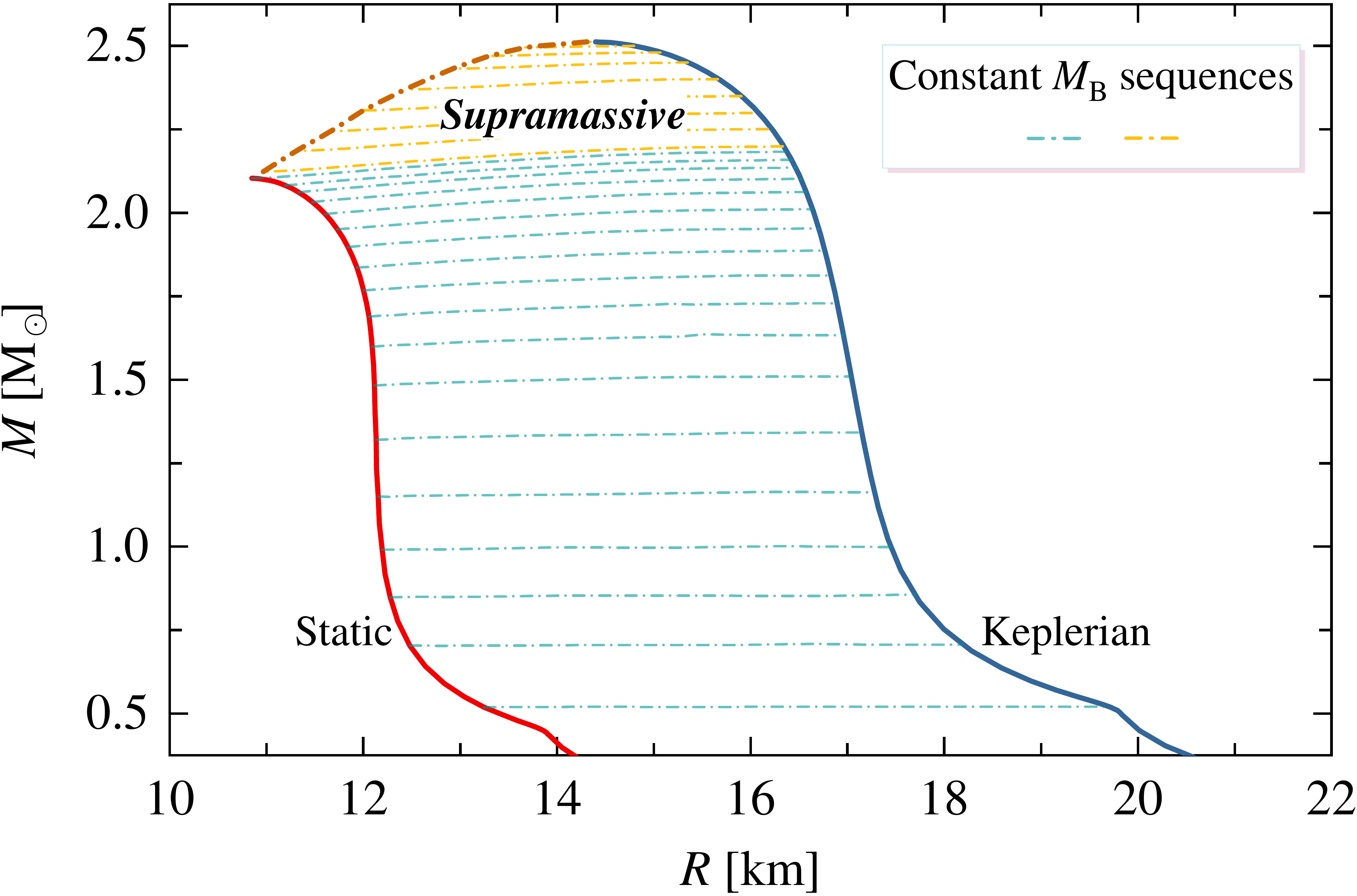}
\caption{Schematic plot showing constant baryonic mass sequences
  calculated with the RHF EoS of NY$\Delta$(c). The supramassive 
  class of stars is sustained by uniform rotation and lacks a 
  static counterpart.}
\label{fig:MR_constant}
\end{figure}
\begin{figure*}[!]
\centering
\includegraphics[width = 0.98\textwidth]{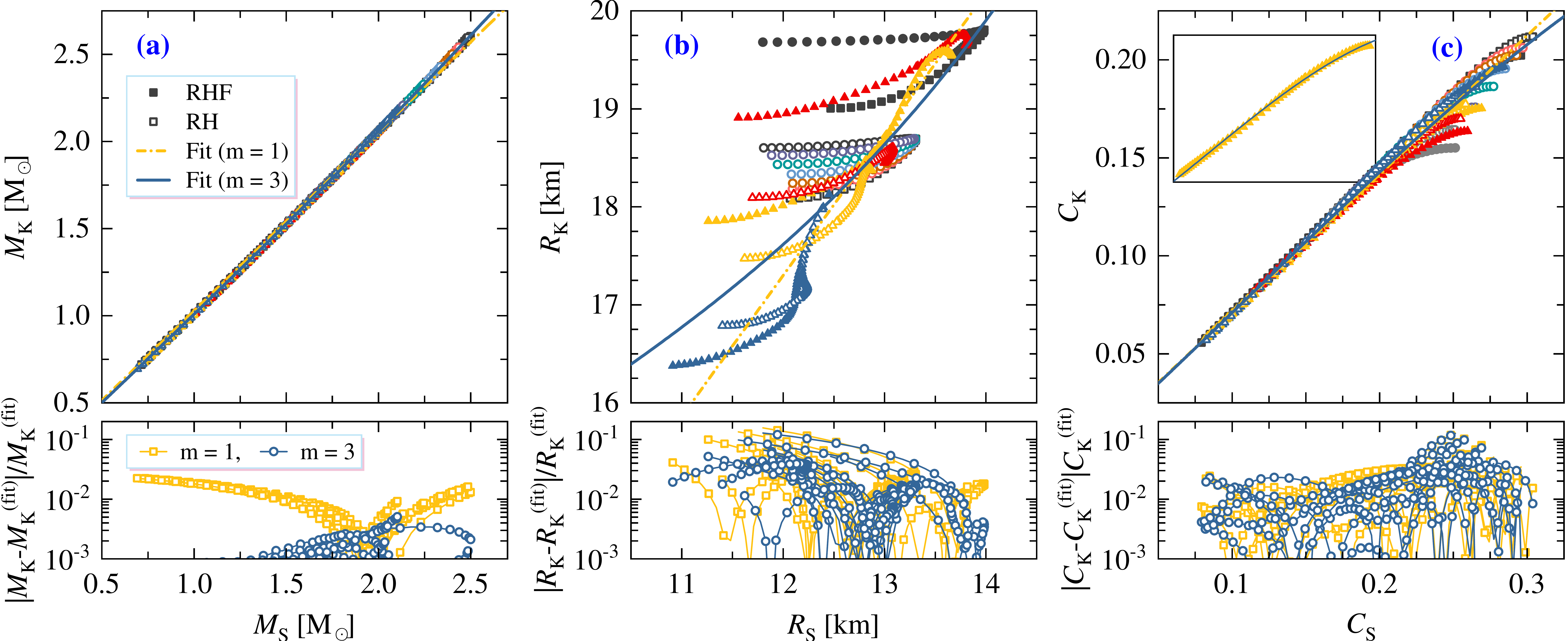}
\caption{The relations between mass, radius, and compactness of
  Keplerian CSs and their static counterparts which have the same
  baryonic mass $M_{\rm B}$, i.e., the sequences are generated by 
  varying $M_{\rm B}$. (Top) Data for various EoS models together 
  with the fitting curves. The result for a single EoS model fitted 
  with a 3rd-order polynomial is illustrated in the inset. (Bottom) 
  Fractional errors between the fitting curves and numerical results.}
\label{fig:MR_constM}
\end{figure*}

In conclusion, it appears that the universal relations discussed 
above do not depend on specific details of the EoS but rather result 
from an overall self-similar behavior of the integrands of global 
quantities. The radial profiles of integrands of quantities observed 
in purely nucleonic EoS models do not exhibit self-similarity in the 
presence of heavy baryons, as revealed by our results. Given that 
these relations hold for EoSs that include heavy baryons, it is evident 
that the mere similarity of the radial profiles is not sufficient to 
guarantee universality.

%-------------------------------------------------------------
\section{Universal relations between the
Keplerian sequence and its non-rotating counterpart}
\label{sec:Results-2}
%-------------------------------------------------------------
In this section, we turn our attention to a particular class of
universal relations, which are related to the integral parameters of
static and maximally (Keplerian) rotating stars. Phenomenologically,
it is relevant to study stellar sequences with constant baryonic mass
$M_{\rm B}$. Evolutionary sequences that lie between the limiting
cases of static stars with mass and radius $M_{\rm S}$ and
$R_{\rm S}$, and Keplerian stars with mass and equatorial radius
$M_{\rm K}$ and $R_{\rm K}$, are often employed to simulate the
spin-down or spin-up of stars under external torques. These torques
can be generated by electromagnetic and gravitational radiation
(resulting in spin-down), or accretion (resulting in spin-up). A
typical case is that of a CS that is born in a supernova and spins
down primarily due to the emission of magnetic-dipole radiation and a
wind of electron-positron pairs, along the evolutionary sequence of
constant baryonic mass.

To illustrate this, Fig.~\ref{fig:MR_constant} shows sequences of
constant baryonic mass ($M_{\rm B}$) calculated using the RHF EoS of
NY$\Delta$(c). It is seen that these sequences represent almost
horizontal lines connecting the Keplerian limiting configuration with
the non-rotating one. Note that the true physical stability may not
terminate at the Keplerian limit as various instabilities may set in
at smaller rotation rates~\citep{Andersson:2001,Bratton:2022}. The
sequences shown can be classified into two categories -- one that does
have a non-rotating stable limit and those with larger masses which do
not. The first category includes stars with a static configuration
mass equal to or less than the maximum mass. The second category
comprises stars that do not have a non-rotating member and are known
as supramassive CSs. In this category, all the stars are unstable 
and terminate in a black hole beyond the maximum star limit for any 
fixed rotation rate.

%----------------------------------------------
\subsection{Relating mass, radius, and compactness 
of static and Keplerian sequences} 
%---------------------------------------------- 
%
\begin{table*}[!]	
\centering
\caption{Estimated coefficients of the relations between the 
  masses, radii, and compactness of Keplerian CSs and their 
  static counterparts, for constant baryonic mass sequences. 
  The corresponding reduced chi-squared ($\chi^2_{\rm red}$) 
  values are given in the last column.}
\setlength{\tabcolsep}{20.8pt}
\begin{tabular}{ccrrrcc}		
\hline\hline
$y$   & $x$ & $a_1$ & $a_2$ & $a_3$ & $\chi^2_{\rm red}$ \\ 
\hline	
$M_{\rm K}$ & $M_{\rm S}$ & $ 1.02816\times 10^{0}$ &              ~&
                                        ~& $ 2.35621\times 10^{-4}$ \\
     ~&      ~& $ 9.97523\times 10^{-1}$ & $ 6.52004\times 10^{-3}$ &
                $ 4.79654\times 10^{-3}$ & $ 7.19416\times 10^{-6}$ \\                                                        
$R_{\rm K}$ & $R_{\rm S}$ & $ 1.44171\times 10^{0}$ &              ~&
                                       ~ & $ 2.82323\times 10^{-1}$ \\
     ~&      ~& $ 3.61768\times 10^{0}$  & $-3.12820\times 10^{-1}$ &
                $ 1.11383\times 10^{-2}$ & $ 2.20765\times 10^{-1}$ \\                  
$C_{\rm K}$ & $C_{\rm S}$ & $ 7.07964\times 10^{-1}$&              ~&
                                        ~& $ 1.98714\times 10^{-5}$ \\
     ~&      ~& $ 6.80281\times 10^{-1}$ & $ 4.18057\times 10^{-1}$ &
                $-1.25613\times 10^{0}$  & $ 1.84240\times 10^{-5}$ \\                                                   
\hline\hline       
\end{tabular}
\label{table-Kepler-1}
\end{table*}
\begin{figure*}[!]
\centering
\includegraphics[width = 0.98\textwidth]{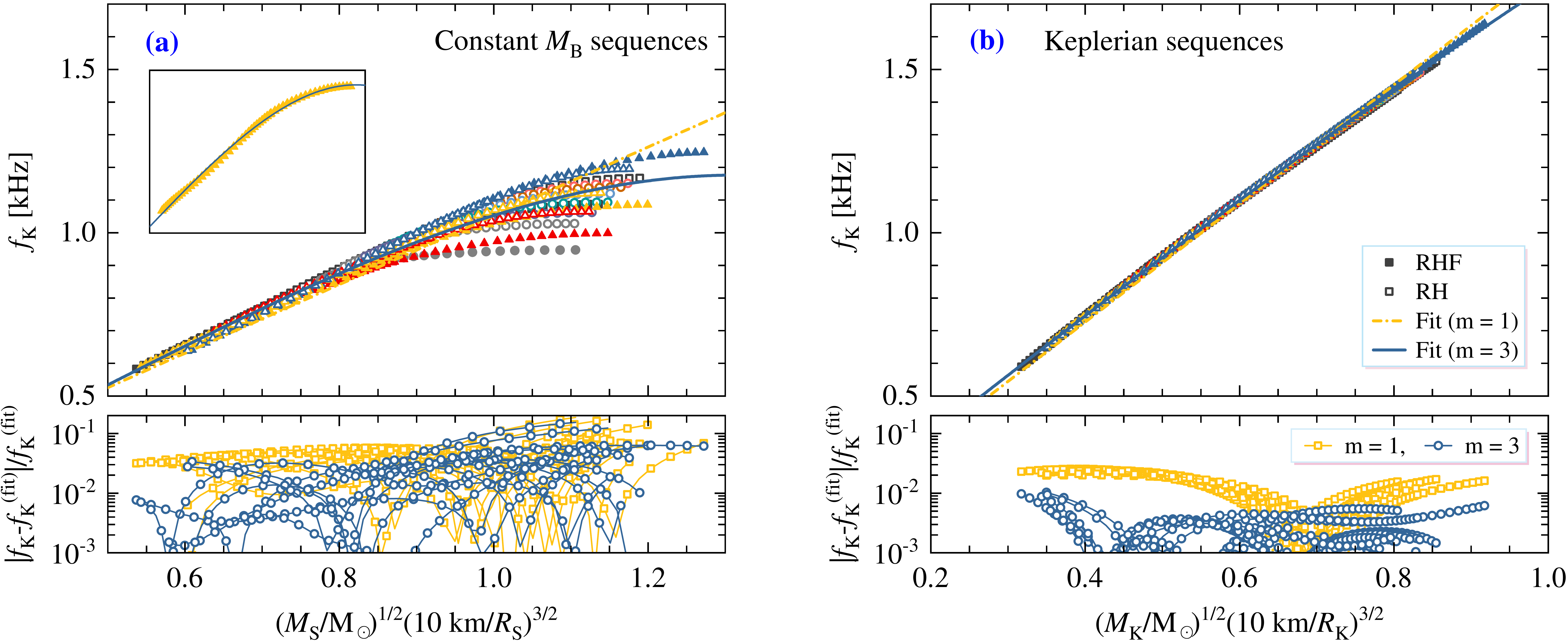}
\caption{(a)~The Keplerian frequencies of CSs as a function of
  $\mathcal{C}_{\rm S}$ [see Eq.~\eqref{eq:C_def}]. The Keplerian
  frequency and static mass and radius values are related by
  assuming that the static and Keplerian CS have same fixed baryonic
  mass, i.e., the sequences are generated by varying $M_{\rm B}$.
  (b)~Same as in (a), but as a function of $\mathcal{C}_{\rm K}$ along
  the Keplerian sequence. (Top) Data from various EoS models and
  fitting curves as shown. The inset shows the result for a single EoS
  model using a 3rd-order fitting polynomial. (Bottom) Fractional
  errors between the fitting curves and numerical results.}
\label{fig:Frequency_kepler}
\end{figure*}

Figure~\ref{fig:MR_constM} depicts the mass, radius, and compactness 
of Keplerian CSs vs the same quantities for their static counterparts
for a fixed baryonic mass of the star $M_{\rm B}$ value for our EoS
collection. In practice, we find for any $M_{\rm S}$ the corresponding 
value of $M_{\rm B}$ and then find the Kepelerian configuration with 
the same value of baryonic mass.

The figure is composed of three panels that correspond to the (a) mass, 
(b) radius, and (c) compactness. These results are then fitted by the 
following polynomial ansatz,
%-------------------------------------------
\begin{align}
y = \sum^m_{n=1} a_n \ x^n, \quad m =1,\,3.
\end{align}
%-------------------------------------------
The relation between $M_{\rm K}$ and $M_{\rm S}$ showed almost linear 
behavior, with deviations of at most a few percent. The relation 
$R_{\rm K}$-$R_{\rm S}$ has a significantly more complicated shape 
and the fits with polynomials provide an accuracy of only the order of 
$10\%$. Per definition, the $C_{\rm K}$-$C_{\rm S}$ relation has a similar 
accuracy. It shows a quasi-linear behavior for small $C_{\rm S}$ 
(large $R_{\rm S}$) values but strong deviations from the linear form 
in the opposite limit of large $C_{\rm S}$ (small $R_{\rm S}$) values. 
Table~\ref{table-Kepler-1} summarizes the fit coefficients for the above 
relations, together with the corresponding values of $\chi^2_{\rm red}$.

%----------------------------------------------
\subsection{Relating Keplerian frequency to static mass and radius} 
%---------------------------------------------- 
%
\begin{table*}[!]	
\centering
\caption{
Estimated coefficients of the relations that involve the 
masses, radii, and frequencies of static and maximally 
(Keplerian) rotating CSs, for constant baryonic mass 
sequences. The corresponding reduced chi-squared 
($\chi^2_{\rm red}$) values are given in the last column.
}
\setlength{\tabcolsep}{20.8pt}
\begin{tabular}{ccrrccc}		
\hline\hline
$y$ & $x$ & $a_1$ & $a_2$ & $a_3$ & $\chi^2_{\rm red}$ \\ 
\hline
$f_{\rm K}$ &    $\mathcal{C}_{\rm S}$    & 
                 $ 1.05183\times 10^{0}$ &                         ~&
                                        ~& $ 2.55674\times 10^{-3}$ \\
     ~&                                                            ~& 
                 $ 7.95060\times 10^{-1}$& $ 8.34800\times 10^{-1}$ &
                 $-5.77261\times 10^{-1}$& $ 1.30980\times 10^{-3}$ \\  
$f_{\rm K}$ &    $\mathcal{C}_{\rm K}$    & 
                 $ 1.81576\times 10^{0}$ &                         ~&
                                        ~& $ 1.56677\times 10^{-4}$ \\
     ~&                                                            ~& 
                 $ 1.91746\times 10^{0}$ & $-1.19466\times 10^{-1}$ &
                 $-3.83918\times 10^{-2}$& $ 9.29333\times 10^{-6}$ \\    	                                                   
\hline\hline       
\end{tabular}
\label{table-Kepler-2}
\end{table*}
\begin{figure*}[tb]
\centering
\includegraphics[width = 0.98\textwidth]{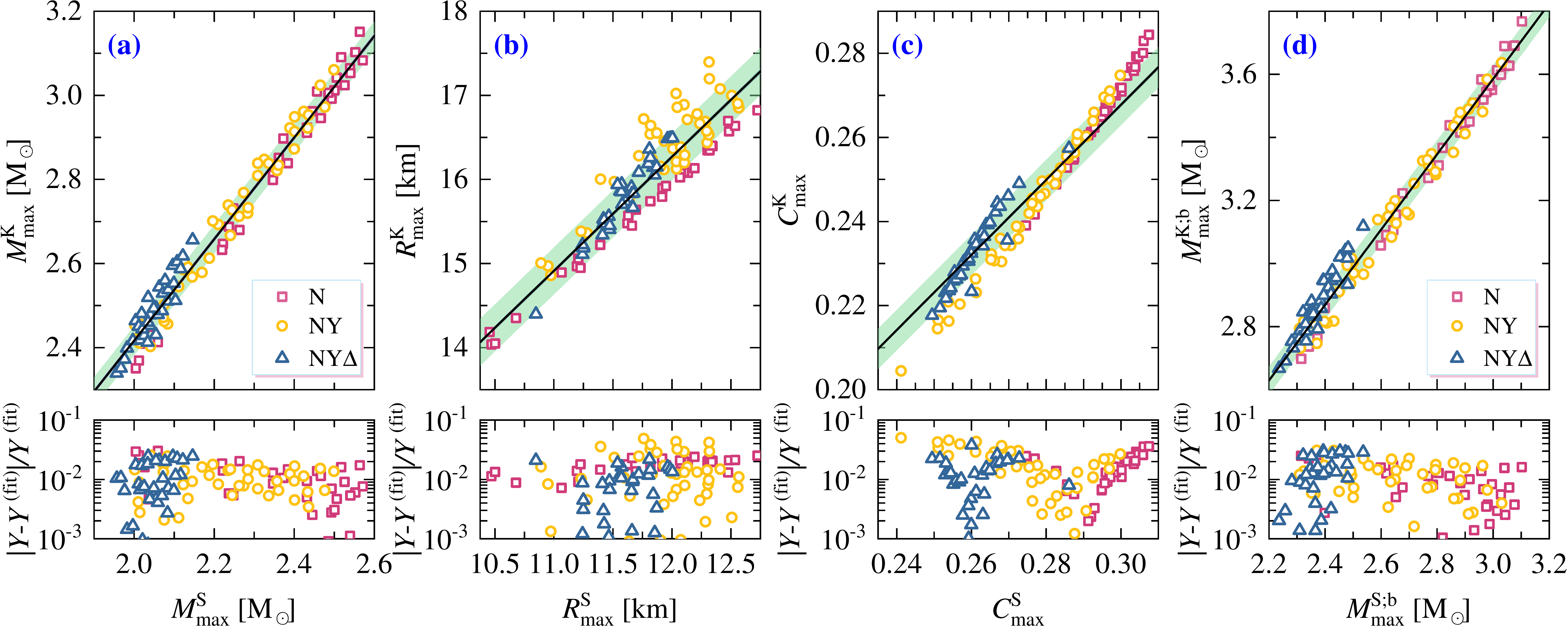}
\caption{The correlations between the gravitational (a) and 
baryonic (d) masses, radii (b), and compactness (c) of maximum-mass 
Keplerian and static stellar configurations. (Top) For each 
quantity the correlation for any  EoS considered is represented by 
a dot and is approximated by a fitting line. The bands depict the 
68.3\% prediction intervals of the linear regressions. (Bottom) 
Fractional differences between the numerical results and the fitting 
lines.}
\label{fig:Mass_radius}
\end{figure*}

We now study the relation between the Keplerian frequency $f_{\rm K}$ 
of a CS and the gravitational mass and radius of the associated static 
star with same baryonic mass $M_B$. In the rigid-body Newtonian limit, 
the Keplerian frequency takes a very simple form, which originates from 
the balance between gravitational and centrifugal forces at the object's 
equator. The Keplerian frequency of a Newtonian sphere with mass $M$ 
and radius $R$ is given by~\citep{Lattimer:2004}
%------------------------------------
\begin{align}\label{Eq:FK_static}	
f^{\rm (N)}_{\rm K} = &\frac{1}{2\pi}\sqrt{\frac{GM}{R^3}} 
		            =  1.8335\left(\frac{M}{M_{\odot}}\right)^{1/2}
		               \left(\frac{10~\mathrm{km}}{R}\right)^{3/2} \mathrm{~kHz}.
\end{align}
%------------------------------------
Considering deformation and the dragging of local inertial frames
within the general theory of relativity~\citep{Haensel:2009}, the
Keplerian frequency exhibits a complicated dependence on the global
structure of a CS. To obtain the Keplerian frequency for a given EoS, 
it is generally necessary to calculate the equilibrium configurations 
within a self-consistent numerical framework, as discussed in 
Appendix~\ref{sec:Models}. Formula~\eqref{Eq:FK_static} is utilized 
to evaluate the general relativistic Keplerian frequency with the 
formula
%------------------------------------
\begin{align}\label{Eq:FK_kepler}	
f_{\rm K}= a \left(\frac{M_\tau}{M_{\odot}}\right)^{1/2}
\left(\frac{10 \mathrm{~km}}{R_\tau}\right)^{3/2}\mathrm{~kHz}
= a\ \mathcal{C}_{\tau} \mathrm{~kHz},
\end{align}
%------------------------------------
where
%------------------------------------
\begin{align}\label{eq:C_def}
\mathcal{C}_{\tau} = \left(\frac{M_\tau}{M_{\odot}}\right)^{1/2}
                     \left(\frac{10~\mathrm{km}}{R_\tau}\right)^{3/2},
\end{align}
%------------------------------------
and $\tau$ denotes the configuration type, where $\tau = \rm S$ 
for static configurations and $\tau = \rm K$ for Keplerian ones.

In Fig.~\ref{fig:Frequency_kepler}, we show the Keplerian frequency 
$f_{\rm K}$ plotted against the corresponding parameters of static 
stars, namely 
$\left({M_{\rm S}}/{M_{\odot}}\right)^{1/2}\left({10\mathrm{~km}}/{R_{\rm S}}\right)^{3/2}$ 
for constant baryonic mass sequences in panel (a), and against themselves, 
$\left({M_{\rm K}}/{M_{\odot}}\right)^{1/2}\left({10\mathrm{~km}}/{R_{\rm K}}\right)^{3/2}$, 
in panel (b). We perform linear fits and 3rd-order extensions for each 
correlation according to
%------------------------------------
\begin{align}	
f_{\rm K}= \sum^m_{n=1} a_n \ (\mathcal{C}_{\tau})^n, \quad m = 1,\,3.
\end{align}
%------------------------------------
Based on Fig.~\ref{fig:Frequency_kepler}, it is apparent that a 
linear correlation is inadequate to fully capture the data, 
suggesting that a more complex or higher-order formula is necessary.
In fact, as shown in the inset of Fig.~\ref{fig:Frequency_kepler}, 
a 3rd-order polynomial could well describe $f_{\rm K}$ for constant 
baryonic mass sequences, computed for a given EoS. We thus perform 
for each relation a 3rd-order polynomial fit.

As depicted in Fig.~\ref{fig:Frequency_kepler}, fitting the data with a 
3rd-order polynomial provides an accuracy of 5\% in assessing the Keplerian 
frequency of a CS using its static counterpart for constant baryonic mass 
sequences up to $0.9\,M^{\rm S}_{\rm max}$, where $M^{\rm S}_{\rm max}$ is 
the maximum mass of the non-rotating (static) configuration. Finally, 
the accuracy reaches 1\% if one writes the Keplerian frequency of the 
star in terms of its mass and radius. The coefficients for each fitting
curve are summarized in Table~\ref{table-Kepler-2}.

%----------------------------------------------
\subsection{Relations between gross properties of 
maximum-mass static and Keplerian configurations}
%---------------------------------------------- 
%
\begin{figure*}[tb]
\centering
\includegraphics[width = 0.98\textwidth]{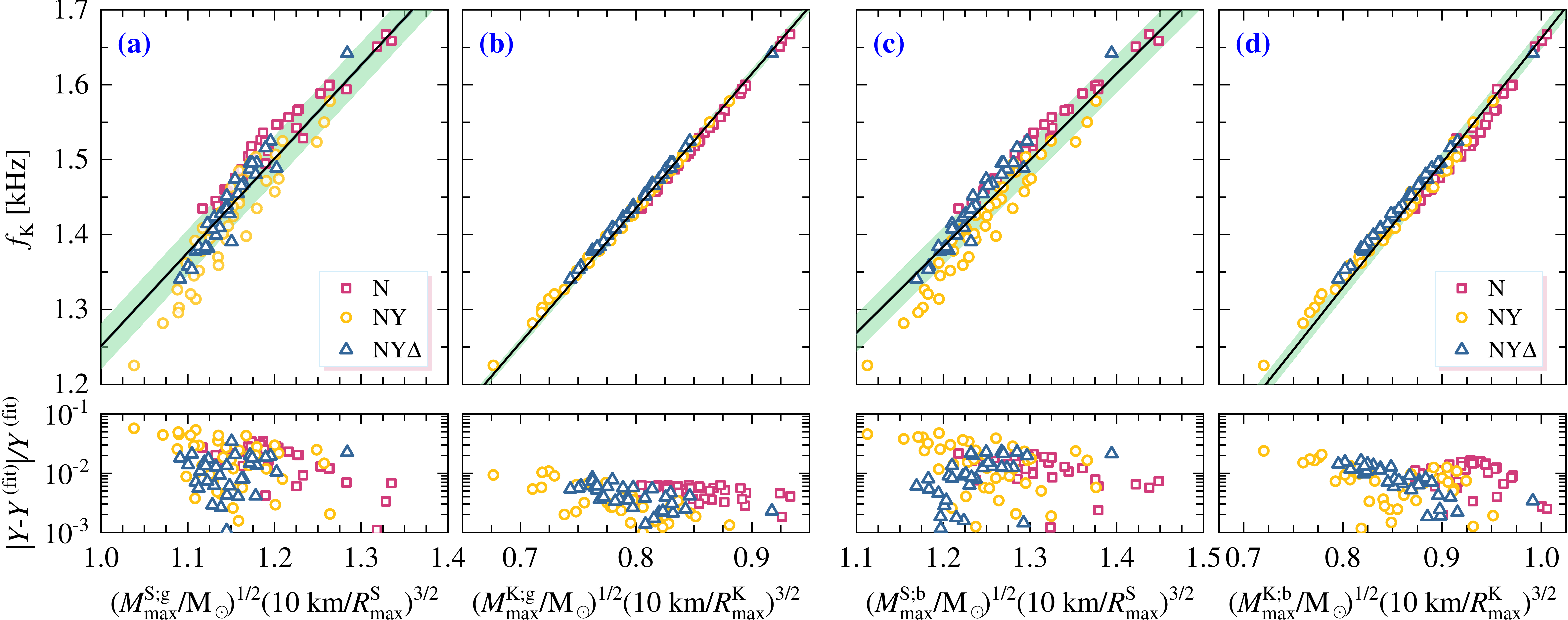}
\caption{The correlation between the Keplerian frequency of 
  maximally rotating CS and $\mathcal{C}^{\tau}_{\rm max}$, 
  see Eq.~\eqref{Eq:C_definition}, where panels (a) and (b) 
  use gravitational static and Keplerian masses respectively, 
  whereas (c) and (d) their baryonic counterparts instead. 
  (Top) For each quantity the correlation for any EoS considered 
  is represented by a dot and is approximated by a fitting line. 
  The bands depict the 68.3\% prediction intervals of the 
  linear regressions. (Bottom) Fractional differences between 
  the numerical results and fitting lines.}
\label{fig:Frequency_mass}
\end{figure*}

As a follow-up to the previous subsection, we next will examine the
relationship between gross properties (gravitational and baryonic mass,
radius, and compactness) of the maximum-mass  static and Keplerian
configurations. Since for each EoS model, we have one data the point 
relating these quantities to the static and maximally rotating 
configurations, it is necessary to incorporate additional EoS
models to quantify and validate the universal relations. Thus, we
consider (i) Nucleonic CDF models proposed in Ref.~\citep{Lijj:2020}, 
which are characterized by two nuclear matter parameters: the 
skewness $Q_{\rm sat}$ and the slope of the symmetry energy 
$L_{\rm sym}$. These models cover 
$2.0 \lesssim M_{\rm max} \lesssim 2.5\,M_{\odot}$ and
$12 \lesssim R_{1.4} \lesssim 14.5$~km. (ii) Hyperonic CDF models
proposed in Ref.~\citep{Lijj:2022}, which are constructed by varying
the values of parameters $L_{\rm sym}$ and $Q_{\rm sat}$ in the
nucleonic sector and the couplings of hyperons in the SU(3) symmetric
model. These models predict
$2.0 \lesssim M_{\rm max} \lesssim 2.5\,M_{\odot}$ and
$12.5 \lesssim R_{1.4} \lesssim 14.5$~km. (iii) $\Delta$-admixed
hyperonic CDF models for the EoS proposed in Ref.~\citep{Lijj:2020},
which are constructed by varying the values of $L_{\rm sym}$ and
$Q_{\rm sat}$ in the nucleonic sector and varying the strength of the
$\Delta$ potential, while setting the couplings of hyperons according
to the SU(6) symmetric model. The latter models lead to
$2.0 \lesssim M_{\rm max} \lesssim 2.2\,M_{\odot}$ and
$12 \lesssim R_{1.4} \lesssim 13.5$~km.
 
The correlations between the gross properties (baryonic and
gravitational masses, radius, and compactness) of maximum-mass static
and Keplerian configurations for each EoS model in our collection are
shown in Fig.~\ref{fig:Mass_radius}. We perform for each correlation 
a linear fit,
%------------------------------------
\begin{align}
y = a\,x,
\end{align}
%------------------------------------
where the values of $x$ and $y$ correspond to quantities in the 
static and maximally rotating configurations, respectively, and 
are used to determine the linear fit for each correlation. 
The coefficient and its standard error related to each fit 
(a linear regression without an intercept term) can be found in 
Table~\ref{table-Kepler-3}. The uncertainty bounds that correspond to 
the standard deviation of the data from this regression coefficient 
are also estimated.

\begin{table}[b]	
\centering
\caption{
The values of the coefficient $a$ for the linear relation between 
the parameters (maximum mass, radius, and compactness) of maximally 
rotating (Keplerian) and static stars. The values quoted in the 
parentheses refer to the standard error of this coefficient. The 
upper and lower bounds (error bars) quoted correspond to the standard 
deviation of the data from this regression coefficient.}
\setlength{\tabcolsep}{5.8pt}
\begin{tabular}{cccc}		
\hline\hline
$y$ & $x$ & $a$ & $\chi^2_{\rm red}$ \\ 
\hline	
$M^{\rm K}_{\rm max}$   & $M^{\rm S}_{\rm max}$   & $1.20815\,(0.00147)\pm 0.01560$ & $1.15397\times 10^{-3}$ \\                                          
$R^{\rm K}_{\rm max}$   & $R^{\rm S}_{\rm max}$   & $1.35578\,(0.00225)\pm 0.02305$ & $7.62623\times 10^{-2}$ \\              
$C^{\rm K}_{\rm max}$   & $C^{\rm S}_{\rm max}$   & $0.89251\,(0.00164)\pm 0.01728$ & $2.25251\times 10^{-5}$ \\ 
$M^{\rm K;b}_{\rm max}$ & $M^{\rm S;b}_{\rm max}$ & $1.19522\,(0.00161)\pm 0.01725$ & $1.91463\times 10^{-3}$ \\                                           
\hline\hline       
\end{tabular}
\label{table-Kepler-3}
\end{table}
\begin{table}[b]	
\centering
\caption{
The values of the coefficient $a$ for the Keplerian frequency 
of maximally rotating stars in terms of their masses and radii, 
as well as those of the maximum-mass static stars. The values 
quoted in the parentheses refer to the standard error of this 
coefficient. The upper and lower bounds (error bars) quoted 
correspond to the standard deviation of the data from this 
regression coefficient.}
\setlength{\tabcolsep}{7.8pt}
\begin{tabular}{cccc}		
\hline\hline
$y$ & $x$ & $a$ & $\chi^2_{\rm red}$ \\  
\hline	
$f_{\rm K}$ & $\mathcal{C}_{\rm max}^{\rm S;g}$ & $1.25106\,(0.00250)\pm 0.02638$ & $9.16179\times 10^{-4}$ \\                                          
$f_{\rm K}$ & $\mathcal{C}_{\rm max}^{\rm S;b}$ & $1.15339\,(0.00190)\pm 0.02010$ & $6.22013\times 10^{-4}$ \\              
$f_{\rm K}$ & $\mathcal{C}_{\rm max}^{\rm K;g}$ & $1.79396\,(0.00078)\pm 0.00824$ & $4.33169\times 10^{-45}$ \\ 
$f_{\rm K}$ & $\mathcal{C}_{\rm max}^{\rm K;b}$ & $1.66183\,(0.00154)\pm 0.01632$ & $1.96817\times 10^{-4}$ \\                                        
\hline\hline       
\end{tabular}
\label{table-Kepler-4}
\end{table}

In this scenario, as shown in Fig.~\ref{fig:Mass_radius}, the linear
fit between masses ($M^{\rm S}_{\rm max},\,M^{\rm K}_{\rm max}$) leads
to a nearly perfect agreement which holds better than 3\%. The 
deviations are by a factor of 2 larger for radii
($R^{\rm S}_{\rm max},\,R^{\rm K}_{\rm max}$) and compactnesses
($C^{\rm S}_{\rm max},\,C^{\rm K}_{\rm max}$). In addition, it is
noteworthy that in Fig.~\ref{fig:Mass_radius}~(b), the data calculated
from purely nucleonic EoS models show an almost linear behavior,
whereas those from heavy-baryons admixed EoS models are distributed
somewhat randomly.

We also show in Fig.~\ref{fig:Mass_radius}~(d) the relationship 
between baryonic masses, where again, we observe an almost linear 
relation between static and Keplerian stars. This suggests the 
existence of a relation between the gravitational mass and baryonic 
mass for CSs in both static and Keplerian configurations.

Finally, we examine the relationships between the Keplerian frequency 
and the gross properties of both static and maximally rotating CSs, 
including the maximum gravitational mass, baryonic mass, and corresponding 
radius~\citep{Lattimer:2004,Haensel:2009,Koliogiannis:2020,Konstantinou:2022}. 
As in Eq.~\eqref{Eq:FK_kepler}, we use the formula
%------------------------------------
\begin{align}	
f_{\rm K} = a\ \mathcal{C}^{\tau}_{\rm max} ~\mathrm{~kHz},
\end{align}
%------------------------------------
where
%------------------------------------
\begin{align}\label{Eq:C_definition}
\mathcal{C}^{\tau}_{\rm max}
  = \left(\frac{M^\tau_{\rm max}}{M_{\odot}}\right)^{1/2}
                      \left(\frac{10~\mathrm{km}}{R^\tau_{\rm max}}\right)^{3/2}.
\end{align}
%------------------------------------
The symbol $\tau$ is used to indicate the form of the configuration, 
where $\tau$ can take on the values $\rm {S; g}$ and $\rm {S; b}$ for 
static configurations with gravitational and baryonic masses, respectively, 
and $\rm {K; g}$ and $\rm {K; b}$ for Keplerian configurations.

Our results are presented in Fig.~\ref{fig:Frequency_mass}. The
maximum rotation frequency is an increasing function of the softness
of an EoS, while the maximum mass decreases. Consequently, the data
predicted by EoSs featuring heavy baryons is located mainly at the
bottom-left part of each figure. As expected, we find a strong
correlation between the Keplerian frequency $f_{\rm K}$ and the
mass-radius combination $\mathcal{C}^{\rm K}_{\rm max}$ of the
rotating configuration defined in Eq.~\eqref{Eq:C_definition}; this
correlation holds an accuracy better than 2\%. The mass-radius combination
$\mathcal{C}^{\rm S}_{\rm max}$ of a static configuration could still
be used to assess the Keplerian frequency $f_{\rm K}$ with an accuracy up
to 6\%. The coefficients for each fitting are summarized in
Table~\ref{table-Kepler-4}.

%-------------------------------------------------------------
\section{Conclusions}
\label{sec:Conclusions}
%-------------------------------------------------------------
In this work, we investigated universal relations for CSs containing 
heavy baryons. We constructed EoS models of dense matter with hyperons 
and $\Delta$-resonances within the covariant density functional theory.
To construct these models, the couplings in the nucleonic sector were 
calibrated by nuclear phenomenology. The couplings of heavy baryons in 
the scalar-meson sector were determined by fitting their potentials at 
symmetric nuclear matter. The vector meson couplings were determined 
using the spin-flavor symmetries of the quark model and its breaking. 
The resulting EoS models are constructed to be consistent with available
constraints from nuclear physics experiments and observations of CSs, 
specifically, radii and masses inferred by the modeling of the NICER 
observations.  The models predict maximum-mass stars and corresponding 
radius values that are within the ranges of
$2.0 \lesssim M_{\rm max} \lesssim 2.5 M_{\odot}$ and
$11.5 \lesssim R_{1.4} \lesssim 14.5$~km.

We first studied the measurable global properties of CSs including
mass, radius, tidal deformability, moment of inertia, and quadrupole
moment for isolated stars.  We have demonstrated that the inclusion of
heavy baryons in CSs does not affect the universal properties of the
$\bar{I}$-$\Lambda$-$\bar{Q}$ relations for static configurations 
(within the slow-rotation approximation) or the $\bar{I}$-$C$-$\bar{Q}$ 
relations for constant spin sequences, as we move from static to 
rapidly rotating configurations. The former relations are found to be 
accurate up to 1\%, while the latter ones have a larger relative error 
of the order of 8\% (except for the $\bar{I}$-$\bar{Q}$ pair). We further 
argue that the $\bar{I}$-$C$-$\bar{Q}$ relation may hold for maximally 
rotating (Keplerian) configurations since the maximum spin parameter
$\chi_{\rm max} = 0.68\pm0.04$ remains constant within 6\% for stars
within the mass range of interest.  The radial dependence of the
integrands entering the calculations of mass, the moment of inertia,
quadrupole moment, and tidal deformability in the Newtonian limit were
examined. It was shown that the presence of heavy baryons breaks the
similarity of the radial profiles, rendering the physical origin of
these relations more complex.

We next investigated the correlations between the properties of
maximally rotating (Keplerian) stars and their non-rotating
counterparts for sequences with the same baryonic mass. It was found
that for sequences with constant baryonic mass, a remarkably tight
correlation exists between their masses, with an accuracy of better
than 0.5\%.  The Keplerian frequency of a maximally rotating star can
be estimated with an accuracy of approximately 10\% and 2\% using the
mass and radius of both the static and rotating stellar configurations, 
respectively. Correlations between the global properties of static and 
rotating maximum-mass configurations were also studied. We found that 
the presence of a significant number of heavy baryons results in slight 
variations in these relations compared to those obtained for CSs 
composed solely of nucleonic matter.

In conclusion, this work presents a study on the universal relations
for CSs that contain heavy baryons. Our derived relations are updated 
versions of those already present in the literature as applied to EoS 
collection with a focus on heavy baryon degrees of freedom. The obtained 
relations can be used to make EoS-insensitive estimates of CS properties 
using, for example, the GW data. Future astronomical observations in 
combination with these relations can be used to improve and expand our 
understanding of the EoS of dense matter.

\section*{Acknowledgements}
J.L. is supported by the National Natural Science Foundation of China
(Grant No. 12105232), the Venture \& Innovation Support Program for
Chongqing Overseas Returnees (Grant No. CX2021007), and by the
``Fundamental Research Funds for the Central Universities" (Grant
No. SWU-020021). A.S. is funded by Deutsche Forschungsgemeinschaft
Grant No. SE 1836/5-2 and the Polish NCN Grant No. 2020/37/B/ST9/01937
at Wrocław University. F.W. acknowledges support provided by the
U.S. National Science Foundation under Grant PHY-2012152.

\appendix
%----------------------------------------------
\section{Equilibrium models for CSs}
\label{sec:Models}
%----------------------------------------------
In this appendix, we briefly specify the relevant equations and
methods that are used to compute the various integral parameters of
static and rotating stars within the general theory of relativity.
The stellar matter is assumed to be a perfect fluid whose
energy-momentum tensor can be written as
%---------------------------
\begin{align}
T_{\mu\nu}=(\varepsilon+P)u_{\mu}u_{\nu}+Pg_{\mu\nu},
\end{align}
%---------------------------
where $\varepsilon$ and $P$ are the energy density and pressure 
of matter, respectively. The quantity $u_{\mu}$ represents the 
four-velocity of the matter and $g_{\mu\nu}$ is the metric tensor.

%----------------------------------------------
\subsection{Static models}
%----------------------------------------------
In the case of static stars, the metric is spherical
symmetrical and is given by
%---------------------------
\begin{align}
ds^2 = -e^{2\nu(r)}dt^2+e^{2\lambda(r)}dr^2+r^2(d\theta^2+\sin^2\!\theta \, d\varphi^{2}),
\end{align}
%---------------------------
where $\nu(r)$ and $\lambda(r)$ are metric functions 
which depend on the radial coordinate $r$.

The static solutions of Einstein's equations are given by the 
Tolman–Oppenheimer–Volkoff (TOV)
equations~\citep{Tolman:1939,Oppenheimer:1939},
%---------------------------
\begin{subequations}\label{tov_equation}
\begin{align}
\label{eq:tov_equation1}
\frac{d m(r)}{d r} & = 
4 \pi r^{2} \varepsilon, \\ 
\label{eq:tov_equation2}
\frac{d P(r)}{\mathrm{d} r} & = 
- \frac{(\varepsilon+ P)(m + 4\pi r^3 P)}{r^2\left(1-\frac{2 m}{r}\right)}, 
\end{align}
\end{subequations}
%---------------------------
where $m(r)$ is the mass enclosed in a mass shell at distance $r$ 
from the center of the star. Equations~\eqref{eq:tov_equation1} and
\eqref{eq:tov_equation2} are numerically integrated over the radial
coordinate $r$ from $r=0$ to $r=R$, where radius $R$ is the radial
distance at which the pressure $P$ becomes zero.

The tidal deformability, $\lambda$, determines how easily an object
can be deformed due to an external tidal
field~\citep{Hinderer:2007,Binnington:2009}. It is given via the
dimensionless tidal Love number $k_2$ and the star’s radius, $R$, 
as $\lambda=2/3k_2R^5$, where
%---------------------------
\begin{align}\label{eq:k2_equation}
k_{2}
=& \frac{8}{5} C^{5}(1-2 C)^{2}\big[2+2 C(y_R-1)-y_R\big] \nonumber \\
& \times\Big\{6C\big[2-y_R+C(5 y_R-8)\big] \nonumber \\
& +4 C^3\big[13-11y_R+C(3 y_R-2)+2 C^2(1+y_R)\big] \nonumber \\
& +3(1-2 C)^2\big[2-y_R+2 C(y_R-1)\big]\ln(1-2 C)\Big\}^{-1}.
\end{align}
%---------------------------
The quantity $C = M/R$ is the dimensionless compactness of 
the star, and $y_R = y(R)$ is extracted from the solution of
%---------------------------
\begin{eqnarray}
\label{eq:tidal_equation}
\frac{\mathrm{d} y(r)}{d r} = -\frac{1}{r}y^2-\frac{1}{r}F_1 y - rF_2,
\end{eqnarray}
%---------------------------
where
%---------------------------
\begin{subequations}
\begin{eqnarray}
F_1 &=& \frac{1-4 \pi r^{2}(\varepsilon-P)}{\left(1- \frac{2m}{r}\right)},\\
F_2 &=& 
\frac{4 \pi}{\left(1- \frac{2m}{r}\right)}
\left(5\varepsilon+9P+\frac{\varepsilon+P}{c^2_s}-\frac{6}{4\pi r^2}\right) \nonumber \\
    & &-4\left[\frac{(m+4 \pi r^3 P)}{r^2\left(1-\frac{2 m}{r}\right)}\right]^{2}.
\end{eqnarray}
\end{subequations}
%---------------------------
Here, $c_s^2=dP/d\varepsilon$ represents the square of the speed of 
sound. Equation~\eqref{eq:tidal_equation} has to be integrated 
simultaneously with the TOV Eqs.~\eqref{eq:tov_equation1}
and~\eqref{eq:tov_equation2} with a boundary value $y(0)=2$. It is
more convenient to work with the dimensionless $\Lambda$, which is
related to the Love number $k_2$ and the compactness parameter through
%---------------------------
\begin{align}
\Lambda = \frac{2}{3}\ k_{2}\ C^{-5}.
\end{align}
%---------------------------
A sequence of stars, each with its mass, radius, tidal
deformability, etc., can be generated by varying the 
central density $\varepsilon_c$ or pressure $P_c$.

%----------------------------------------------
\subsection{Rotating models}
% ----------------------------------------------
Highly accurate numerical methods for computing the properties 
of rotating stars have been
developed~\citep{Komatsu:1989,Cook:1994,Stergioulas:1994,Bonazzola:1993,Ansorg:2003}.
Equilibrium configurations of rotating stars can be computed using the
publicly available RNS~\citep{Stergioulas:1994,Stergioulas:2003} and
LORENE/rotstar~\citep{Bonazzola:1993,Bonazzola:1998} codes.

We employed the RNS
code~\footnote{http://www.gravity.phys.uwm.edu/rns} to obtain the
results for rotating stars reported below. It solves the Einstein
field equations for an axisymmetric and stationary space-time
described by the metric
%-----------------------
\begin{align}
ds^2 = & -e^{\gamma+\rho}dt^2 + e^{2\alpha}(dr^2+r^2d\theta^2) \nonumber \\ 
       & + e^{\gamma-\rho}r^2\sin^2\!\theta \, (d\phi-\omega dt)^2,
\end{align}
%-----------------------
where $\gamma$, $\rho$, $\alpha$, and $\omega$ are metric potentials 
that depend on the radial coordinates $r$ and the polar angle $\theta$.

Using the RNS code, we computed the mass, radius, moment of inertia
and quadrupole moment of rotating star for a specified central
density. Our results take into account the correction for the quadrupole 
moment $Q$ given in Refs.~\citep{Pappas:2012,Yagi:2014}.

Dimensionless quantities for the moment of inertia and the quadrupole
moment can be defined as follows:
%---------------------
\begin{align}
\bar{I} = \frac{I}{M^3} \quad \text{and} \quad \bar{Q} = -\frac{Q}{M^3\chi^2},
\end{align}
%---------------------
where $M$ is the gravitational mass, $\chi= J/M^2$ represents the 
dimensionless spin parameter, and $J$ is the angular momentum. The 
oblate shape of a rotating CS is approximately universal and can be
described by a function of the equatorial compactness
$C_{\rm eq} = M/R_{\rm eq}$ (where $R_{\rm eq}$ is the radius of the
star measured at the equator) and the angular
velocity~\citep{Morsink:2007}. In this work, we will use $C$ to refer
to the equatorial compactness of rotating stars.

Finally, for completeness, we quote the equation for the general
relativistic Keplerian frequency $f_{\rm K}$  for a test particle moving 
along the orbit with semimajor axis $a_{\rm K}$ around a body with 
mass $M$,
% ----------------------------
\begin{eqnarray}
f_{\rm K} = \left(\frac{GM}{a^3_{\rm K}}\right)^{1/2} \left[1 -
  \left(\frac{3GM}{c^2a_{\rm K}}\right)
  \left(1 - \epsilon^2\right)
  \right],
\end{eqnarray}
%----------------------------
where $\epsilon$ is the eccentricity of the orbit. It reduces to
Eq.~\eqref{Eq:FK_kepler} by considering a test particle moving along
the equatorial circumference of the Keplerian star of mass $M_{\rm K}$.

%\bibliography{Urelation_refs}
%

\end{document}